\documentstyle[aps,amssymb]{revtex}

\begin{document}
\author{R. W. Spekkens and J. E. Sipe}
\title{Non-orthogonal preferred projectors for modal interpretations of quantum
mechanics}
\date{March 20, 2000}
\address{Department of Physics, University of Toronto, 60 St. George Street, Toronto,
Ontario, Canada M5S 1A7}
\maketitle

\begin{abstract}
Modal interpretations constitute a particular approach to associating
dynamical variables with physical systems in quantum mechanics. Given the
`quantum logical' constraints that are typically adopted by such
interpretations, only certain sets of variables can be taken to be
simultaneously definite-valued, and only certain sets of values can be
ascribed to these variables at a given time. Moreover, each allowable set of
variables and values can be uniquely specified by a single `preferred'
projector in the Hilbert space associated with the system. In general, the
preferred projector can be one of several possibilities at a given time. In
previous modal interpretations, the different possible preferred projectors
have formed an orthogonal set. This paper investigates the consequences of
adopting a non-orthogonal set. We present three contributions on this issue:
(1) we provide an argument for such non-orthogonality, based on the
assumption that perfectly predictable measurements reveal pre-existing
values of variables, an assumption which has traditionally constituted a
strong motivation for the modal approach; (2) we generalize the existing
framework for modal interpretations to accommodate non-orthogonal preferred
projectors; (3) we present a novel type of modal interpretation wherein the
set of preferred projectors is fixed by a principle of entropy minimization,
and we discuss some of the successes and shortcomings of this proposal.
\end{abstract}

\section{\strut Introduction}

In operational quantum mechanics, theoretical predictions take the form `if
such-and-such a measurement is made after such-and-such a preparation,
such-and-such an outcome will be found{\em \ }with such-and-such a
probability'. In contrast, a {\em realist} interpretation is an attempt to
understand quantum mechanics as making stronger claims of the form
`such-and-such a variable has such-and-such a value with such-and-such a
probability'. The `elements of reality' of Einstein, Podolsky and Rosen\cite
{EPR}, and Bell's `be-ables'\cite{Bell} are two ways of referring to the
variables that possess definite values in a realist interpretation. We will
simply refer to them as the {\em determinate variables}. Since the set of
determinate variables in some sense specifies `what exists', we call it the 
{\em ontology} for the system. The specification of the values of the
determinate variables will be called the {\em value ascription }to the
ontology. Within this approach, one assigns a property to a system by
assigning a value to a determinate variable. Since the ontology and the
value ascription together constitute a complete specification of the
properties of a quantum system, they will jointly be referred to as the {\em %
property ascription.}

Within `orthodox' interpretations of this type, a variable is determinate if
and only if it is associated with an operator for which the state vector is
an eigenstate, and its value is the corresponding eigenvalue. It is also
assumed that a variable $V$ defined on a subsystem is determinate only if $%
V\otimes I$ is determinate on the total system, where $I$ is the identity
operator for the part of the total system that is not included in the
subsystem. According to these rules, the ontologies and the value
ascriptions for all systems are uniquely defined by the state vector.{\em \ }%
It is widely recognized that this view, together with the assumption that
the evolution of the state vector is unitary for all time, leads to the
quantum measurement problem, namely, the failure to ensure the
determinateness of macroscopic variables such as the pointer reading of an
apparatus\cite{BubIQW}.

One approach to the problem is to introduce a non-unitary dynamics for the
state vector into the formalism of the theory (the `collapse' of the state
vector). A different approach is to preserve the unitary dynamics, but to
reject the notion that a variable is determinate only if it has the state
vector as an eigenstate of the associated operator. In the latter type of
approach the property ascription need not be fixed at a given time by the
state vector. Rather, it may be that the state vector describes only the set
of {\em possible} property ascriptions, in which case it describes what is
possible and what, if anything, is necessary. Since the logic of possibility
and necessity is modal logic, realist no-collapse interpretations of this
type have been called {\em modal interpretations }of quantum mechanics\cite
{see Clifton}.

Modal interpretations typically impose many constraints on the form of the
property ascription for a system. Given these constraints, there is always a
unique `most elementary' possessed property defined by a property ascription%
{\em .} We call this the {\em preferred property} for that property
ascription. Since at a given time the property ascription may be one of
several possibilities, each of which define a different preferred property,
there is in general a set of preferred properties associated with a system.
An example may serve to clarify these concepts. Suppose the system is a
digital display on an apparatus. The property ascription for the display may
include such properties as `the digital display shows a number between $1$
and $3$', `the digital display shows a number smaller than $5$', etc., while
the preferred property may be `the digital display shows the number $2$'.
The set of preferred properties may consist of a list of properties each of
the form `the digital display shows the number $k$', but differing in the
value of $k.$ In previous modal interpretations, the preferred properties
have been associated with {\em orthogonal }projectors.

There are three contributions made in this paper. First, we demonstrate that
any modal interpretation which adopts the standard constraints upon the
property ascription and which seeks to satisfy a particular criterion of
faithful measurement must allow for the set of preferred properties to be
associated with non-orthogonal projectors. Second, we introduce a framework
for modal interpretations that incorporates such preferred properties.
Third, we present a novel proposal within this framework wherein the
preferred properties are fixed by a principle of entropy minimization.

The paper is organized as follows. In section 2, we present a review of the
constraints upon the property ascription that are standard among modal
interpretations, and we provide a rigorous definition of the notion of a
preferred property. In section 3, we present the argument that the preferred
properties must be associated with non-orthogonal projectors if one hopes to
explain the outcomes of perfectly predictable measurements in terms of
pre-existing properties of the system under investigation, that is, if
perfectly predictable measurements are to be faithful.{\em \ }The argument
relies on a particular kind of experiment, involving a sequence of two
measurements which have the following critical features: (1) the first
measurement disturbs the state of the system differently for different
outcomes, resulting in the preparation of non-orthogonal states; and (2) the
variable measured by the second device depends on the outcome of the first
measurement in such a way that the outcome of the second measurement is
always perfectly predictable.

In order to accomodate non-orthogonal preferred properties, we require a new
framework for modal interpretations, which is the subject of section 4. We
preserve most of the standard constraints on the property ascription, in
particular constraints involving the functional relations between the values
of variables. However, we show that one must abandon the assumption that
different property ascriptions share a common ontology. We assume Healey's
so-called `weakening condition'\cite{Healey}, and adopt Clifton's rule\cite
{Cliftonproperties} for relating the properties of composites to the
properties of the subsystems of which they are composed. Moreover, we follow
previous authors\cite{BacciagaluppiDickson}\cite{Dicksonplurality} in
requiring that the dynamics of the property ascription be Markovian and
satisfy certain constraints of analyticity, while also reproducing the
standard quantum statistics. Guided by these constraints, we introduce a
framework for modal interpretations that incorporates non-orthogonal
preferred projectors. This constitutes a generalization of the framework
introduced by Bub and Clifton\cite{Bub and Clifton}.

With this framework in hand, we proceed in section 5 to present a novel
proposal for a modal interpretation. We begin by assuming that there is a
distinguished division of the universe into elementary subsystems, or
equivalently a distinguished factorization of the total Hilbert space. A
preferred decomposition of the state vector is singled out by the
minimization of a particular entropic quantity that quantifies the degree of
entanglement of the state vector with respect to the distinguished
factorization. The actual property ascription is assumed to be fixed by a
single element of this decomposition, whose identity evolves by a stochastic
process with specified statistical properties. Within this proposal, we
demonstrate that the quantum measurement problem is avoided for several
models of measurement interactions, and a large class of perfectly
predictable measurements are shown to reveal pre-existing properties. In
section 6, we present our concluding remarks.

\section{The Modal Approach}

\subsection{Review of Constraints}

We begin by considering the notion of a {\em property} of a physical system.
The type of properties in which we are interested are those of the form
`having a value of the variable $V$ in the range $\Delta $'. For every such
property, one can associate an idempotent variable that has value $1$ if the
value of $V$ is in the range $\Delta $ and $0$ if the value lies outside
this range. Whether a property is possessed or not is given by the value of
this idempotent variable; it is possessed if the value is $1,$ and it is not
possessed if the value is $0.$

In classical mechanics, a variable is represented by a function on phase
space, and the possible values of this variable are just the values in the
range of this function. The property of `having a value of the variable $V$
in the range $\Delta $' is associated with the subset of phase space
containing all points for which $V$ is in the range $\Delta $. For instance,
if the system is a one-dimensional harmonic oscillator, the property `having
energy between $E_{1}$ and $E_{2}$' is associated with an elliptical ring in
phase space, while the property `having position between $x_{1}$ and $x_{2}$%
' is associated with a vertical band. Suppose the property $s$ is associated
with a subset $\Omega $ of phase space. The idempotent variable associated
with the property $s$ is the function $F_{\Omega }$ that takes the value 1
for every point in $\Omega $ and $0$ for every point outside $\Omega $.{\em %
\ } If two properties $s$ and $s^{\prime }$ are represented by subsets $%
\Omega $ and $\Omega ^{\prime }$ of phase space, then the disjunction of $s$
and $s^{\prime }$ is represented by the union of $\Omega $ and $\Omega
^{\prime },$ the conjunction of $s$ and $s^{\prime }$ is represented by the
intersection of $\Omega $ and $\Omega ^{\prime },$ and the negation of $s$
is represented by the complement of $\Omega .$ The truth tables appropriate
for conjunction, disjunction and negation in classical logic place the
following constraints on the values of the idempotent variables: 
\begin{eqnarray*}
F_{\Omega \cup \Omega ^{\prime }} &=&F_{\Omega }+F_{\Omega ^{\prime
}}-F_{\Omega }F_{\Omega ^{\prime }} \\
F_{\Omega \cap \Omega ^{\prime }} &=&F_{\Omega }F_{\Omega ^{\prime }} \\
F_{\overline{\Omega }} &=&1-F_{\Omega },
\end{eqnarray*}
where $\cup ,\cap $ and an over-bar respectively denote union, intersection
and complementation of subsets of phase space.

We now consider an approach to realist interpretations of quantum mechanics
that parallels those features of classical theories described above, except
of course that the mathematical structure relevant for the description of a
system is no longer phase space but Hilbert space. This approach has its
origin in the field of quantum logic\cite{quantum logic}, and is adopted by
most modal interpretations. A variable is represented by a Hermitian
operator over the Hilbert space, and its possible values are the eigenvalues
of this operator. Properties are associated with subspaces of Hilbert space.
The idempotent variable associated with a given property is represented by
the projector onto the corresponding subspace. If two properties $s$ and $%
s^{\prime }$ are represented by subspaces ${\cal S}$ and ${\cal S}^{\prime }$%
, then the disjunction of $s$ and $s^{\prime }$ is represented by the linear
span (direct sum) of ${\cal S}$ and ${\cal S}^{\prime }$, the conjunction of 
$s$ and $s^{\prime }$ is represented by the intersection of ${\cal S}$ and $%
{\cal S}^{\prime }$, and the negation of $s$ is represented by the
orthogonal complement of ${\cal S}$. These assumptions will be called the%
{\bf \ }{\em constraints on logical connectives}{\bf .}

For simplicity, we denote both the projector onto the subspace ${\cal S}$
and the associated idempotent variable by $P_{{\cal S}}$, and denote the
value of this idempotent variable by $[P_{{\cal S}}].$ Analogously to the
classical case, we adopt the following constraint on the values of the
idempotent variables:

{\em Functional relation constraint} 
\begin{eqnarray}
\lbrack P_{{\cal S}\oplus {\cal S}^{\prime }}] &=&[P_{{\cal S}}]+[P_{{\cal S}%
^{\prime }}]-[P_{{\cal S}}][P_{{\cal S}^{\prime }}]  \label{ortho 1} \\
\lbrack P_{{\cal S\cap S}^{\prime }}] &=&[P_{{\cal S}}][P_{{\cal S}^{\prime
}}]  \label{ortho 2} \\
\lbrack P_{{\cal S}^{\perp }}] &=&1-[P_{{\cal S}}],  \label{ortho 3}
\end{eqnarray}
where $\oplus ,$ $\cap $ and $^{\perp }$ denote respectively linear span,
intersection, and orthogonal complement.

As it turns out, it is impossible to associate values with all the
projectors in a Hilbert space in a way that is consistent with the
functional relation constraint and the constraints on logical connectives 
\cite{no 2-valued homomorphisms}. The response of modal interpretations is
to associate definite values with only a subset of all the projectors in the
Hilbert space. Thus, in contrast with classical mechanics, only a subset of
all idempotent variables correspond to well-defined properties at any given
time. The projectors that are associated with definite values are labelled 
{\em determinate, }as are the corresponding idempotent variables. The
functional relation constraint is only required to hold among the
determinate projectors.

As regards non-idempotent variables, we adopt the convention that a variable
is denoted by the same symbol as the associated hermitian operator, for
instance $V,$ and that its value is denoted by $[V].$ Moreover, following
other modal interpreters\cite{Clifton motivating}, we adopt the attitude
that if the spectral resolution of a Hermitian operator is $%
V=\sum_{k}v_{k}P_{{\cal S}_{k}},$ where $v_{k}\ne 0$ for all $k,$ and $%
v_{k}\ne v_{k^{\prime }}$ for $k\ne k^{\prime },$ then $V$ is determinate if
and only if all of the projectors in the set $\{P_{{\cal S}_{k}}\}_{k}$ are
determinate, and in this case its value is $[V]=\sum_{k}v_{k}[P_{{\cal S}%
_{k}}]$. We call this the {\em spectral constraint}{\bf . }

It follows from this constraint that the set of determinate idempotent
variables and their values are sufficient to specify the set of all
determinate variables and their values. As noted in the introduction, we
will refer to the set of determinate variables as the {\em ontology}, and
the values of these variables as the {\em value ascription}. The ontology
and the value ascription together define the {\em property ascription.}

We turn now to constraints on the nature of the ontology. It is typically
assumed that logical combinations of well-defined properties are also
well-defined. Thus, if property $s$ is well-defined, then so too should be
the property `not $s$'$.$ In other words, if $P_{{\cal S}}$ is determinate,
then $P_{{\cal S}^{\perp }}$ ($=I-P_{{\cal S}})$ should be determinate as
well. Similarly, if properties $s$ and $s^{\prime }$ are well-defined, so
that the associated projectors $P_{{\cal S}}$ and $P_{{\cal S}^{\prime }}$
are determinate, then the properties `$s$ or $s^{\prime }$' and `$s$ and $%
s^{\prime }$' should also be well-defined, and the associated projectors $P_{%
{\cal S}\oplus {\cal S}^{\prime }}$ and $P_{{\cal S}\cap {\cal S}^{\prime }}$
should be determinate. In summary, we require

{\em Closure constraint}{\bf :} 
\begin{eqnarray*}
&&\text{If }P_{{\cal S}}\in Ont\text{ then }P_{{\cal S}^{\perp }}\in Ont \\
\text{If }\{P_{{\cal S}},P_{{\cal S}^{\prime }}\} &\in &Ont\text{ then }\{P_{%
{\cal S}\oplus {\cal S}^{\prime }},P_{{\cal S}\cap {\cal S}^{\prime }}\}\in
Ont,
\end{eqnarray*}
where $Ont$ denotes the ontology. Assuming that for every system there is at
least one projector that is determinate, it follows from the closure
constraint that the identity operator and the projector onto the null space
are determinate for every system.

We refer to all of the constraints on the ontology and value ascription that
have been presented thus far as the {\em algebraic constraints}.

\subsection{Preferred Projectors}

We now derive a few consequences of the algebraic constraints. For this
purpose, it is useful to introduce a notational convenience: `$P_{{\cal S}%
}>P_{{\cal S}^{\prime }}$' denotes that ${\cal S}^{\prime }$ is a subspace
of ${\cal S},$ and `$P_{{\cal S}}\perp P_{{\cal S}^{\prime }}$' denotes that 
${\cal S}$ and ${\cal S}^{\prime }$ are orthogonal subspaces. First, we show
that in every property ascription there is a projector $P_{{\cal S}}$ that
receives the value $1,$ and for which no projector onto a proper subspace of 
${\cal S}$ receives the value 1; that is, there is a subspace ${\cal S}$
such that 
\begin{equation}
\lbrack P_{{\cal S}}]=1\text{ and there is no subspace }{\cal T}\text{ such
that }P_{{\cal T}}<P_{{\cal S}}\text{ and }[P_{{\cal T}}]=1.
\label{defintion of pref proj}
\end{equation}
Such a subspace always exists since at least one projector, namely the
identity operator, always receives the value 1. Such a subspace is unique
because if ${\cal S}$ and ${\cal S}^{\prime }$ were two distinct subspaces
satisfying this definition, then both would receive the value $1,$ and by
the functional relation constraint their intersection would also receive the
value $1,$ which implies that either ${\cal S}$ or ${\cal S}^{\prime }$ has
a proper subspace that receives the value $1.$ We call the unique projector
satisfying Eq. (\ref{defintion of pref proj}) the {\em preferred projector}
for the property ascription. The property associated with this projector is
also called preferred.

A modal interpretation must account for the fact that a measurement device
may have one of several different properties at the end of a measurement
despite there being a single state vector for the universe. This is
accomplished by assuming that the property ascription is not fixed by the
state vector, but may be one of several possibilities at a given time. There
are two notions of possibility that are adopted by modal interpreters in
this context. In the first, the different possibilities for the final
property ascription to the measurement device are attributed to differences
in the initial properties(possibly hidden) of the system. In the second, the
different possibilities for the final property ascription to the measurement
device arise from an objective stochasticity in the evolution of the
property state\footnote{%
An example of the first approach is the Bub-Clifton interpretation when the
preferred variable has a continuous unbounded spectrum, and the dynamics is
given by a guidance equation analogous to the one used in Bohmian mechanics.
An example of the second approach is the Bub-Clifton interpretation when one
adopts a different guidance equation, or when the preferred variable has a
discrete spectrum. See, for instance, section 5.2 of Ref. \cite{BubIQW}.}.
Although we adopt the latter view in subsequent sections, for the present it
suffices to note that in all modal interpretations one associates with a
system, at every time, one out of a set of several possible property
ascriptions. Since each such property ascription defines a preferred
projector, there is in general a set of preferred projectors associated with
a system at a given time. We now consider the relation between the elements
of this set.

We begin with a definition: two properties are said to be {\em mutually
exclusive} if their conjunction is a contradiction. This is stronger than
simply being distinct, as is illustrated by the properties `red' and `red or
blue', which are distinct, but not mutually exclusive. We also use the term
`mutually exclusive' to describe two property ascriptions if there is a
property obtaining in one that is mutually exclusive to a property obtaining
in the other. We assume that for all systems at all times, the different
possible property ascriptions are mutually exclusive. This assumption can be
recast as a constraint upon the set of preferred projectors. Recalling the
constraints on logical connectives,{\em \ }for two property ascriptions to
be mutually exclusive there must be two projectors $P_{{\cal R}}$ and $P_{%
{\cal R}^{\prime }}$ such that $[P_{{\cal R}}]=1$ in the first property
ascription and $[P_{{\cal R}^{\prime }}]=1$ in the second$,$ but such that
the intersection of ${\cal R}$ and ${\cal R}^{\prime }$ is the null space.
By definition, the preferred projector for the first property ascription$,$
call it $P_{{\cal S}},$ must be such that $P_{{\cal S}}\le P_{{\cal R}},$
and the preferred projector for the second$,$ call it $P_{{\cal S}^{\prime
}},$ must be such that $P_{{\cal S}^{\prime }}\le P_{{\cal R}^{\prime }},$
from which it follows that the intersection of ${\cal S}$ and ${\cal S}%
^{\prime }$ must also be the null space. Thus we conclude that the preferred
projectors for mutually exclusive property ascriptions are associated with
subspaces whose intersection is the null space. Note that we have {\em not }%
concluded that the preferred projectors are {\em orthogonal}. Indeed, the
possibility of preferred projectors that are non-orthogonal will be the
focus of much of this paper.

\section{Non-orthogonal Preferred Projectors}

\subsection{The Faithfulness Criterion}

One can debate the merits of assuming orthogonal preferred projectors in the
context of a macroscopic system such as the pointer on a measurement device.
On the one hand, the distinguishability of the different physical states of
a pointer suggest that they must be associated with orthogonal projectors;
on the other hand, the requirement of quantum-classical correspondence
suggests that the alternative positions of a pointer should be associated
with projectors onto a set of coherent states, or some similar over-complete
and non-orthogonal basis. We do not take a stand on this issue here. We do
provide an argument for adopting a non-orthogonal set of preferred
projectors, but it appeals to the properties that should be assigned to {\em %
microscopic} rather than macroscopic systems. In particular, we consider a
microscopic system that is the object of investigation in a quantum
measurement.

Since operational quantum mechanics only makes reference to the properties
of macroscopic preparation and measurement devices, the requirement of
agreement with the operational theory does not by itself constrain the
properties that are assigned to the microscopic systems under investigation.
But if the properties of such microscopic systems are to play any
explanatory role in the theory, then one would expect their role to be in
determining the outcomes of measurements upon them. In particular, we
consider the following criterion for assigning determinate status to a
variable:

\begin{description}
\item[Faithfulness Criterion]  \ If it can be predicted with probability $1$
that a measurement of the variable $V$ will yield the result $v$, then
immediately prior to the measurement the variable $V$ is determinate with
value $v.$
\end{description}

The motivation for adopting this criterion is presented in the next
subsection. It is nonetheless worth emphasizing at this point its importance
in this paper: the particular form of the framework for modal
interpretations that is presented in section 4 and the particular proposal
presented in section 5 are both to a large extent attempts to satisfy the
faithfulness criterion.

This criterion is applicable to experiments involving a sequence of
measurements, the first of which may be considered a preparation. Within the
context of operational quantum mechanics, we consider a sequence of two
measurements, associated with distinct Hermitian operators, $V$ and $%
V^{\prime },$ both belonging to the Hilbert space ${\cal H}^{S}.$ For
simplicity, we assume that these operators have non-degenerate eigenvalues,
denoted by $\{v_{k}\}_{k=1}^{m}$ and $\{v_{j}^{\prime }\}_{j=1}^{m}$
respectively, and eigenvectors denoted by $\{\left| \varphi
_{k}\right\rangle \}_{k=1}^{m}$ and $\{\left| \varphi _{j}^{\prime
}\right\rangle \}_{j=1}^{m}$ respectively. In order to predict the outcome
of the second measurement, it is necessary to also specify how, if at all,
the first measurement disturbs the state. Suppose then that upon obtaining
outcome $k$ for the first measurement, the state $\left| \tilde{\varphi}%
_{k}\right\rangle $ is prepared, and that the set of vectors $\{\left| 
\tilde{\varphi}_{k}\right\rangle \}_{k=1}^{m},$ although normalized and
non-collinear, are non-orthogonal. Since it follows that $\left| \tilde{%
\varphi}_{k}\right\rangle \ne \left| \varphi _{k}\right\rangle $ for one or
more values of $k,$ we say that the measurement is {\em disturbing}. For
simplicity, the second measurement is assumed to be non-disturbing, and the
two measurements are assumed to be immediately consecutive. Finally, we take
the preparation procedure that precedes the first measurement to be
associated with a state vector $\sum_{k}c_{k}\left| \varphi
_{k}\right\rangle ,$ where $\sum_{k}\left| c_{k}\right| ^{2}=1,$ and $%
c_{k}\ne 0$ for all $k.$

Operational quantum mechanics predicts, via the generalized Born rule, that
the probability of the second apparatus indicating outcome $j$ given that
the first apparatus indicates outcome $k$ is 
\[
\text{Prob}(j|k)=\left| \left\langle \tilde{\varphi}_{k}|\varphi
_{j}^{\prime }\right\rangle \right| ^{2}. 
\]
It follows that in order for the outcome of the second measurement to be
predictable with probability 1 given the outcome of the first measurement,
it must be the case that $\left\langle \tilde{\varphi}_{k}|\varphi
_{j}^{\prime }\right\rangle =1$ for some values of $k$ and $j.$ Hence, the
variable $V^{\prime }$ measured by the second apparatus must have at least
one of the states in the set $\{\left| \tilde{\varphi}_{k}\right\rangle
\}_{k=1}^{m}$ as an eigenstate. We define a set of variables $%
\{V_{(k)}\}_{k=1}^{m},$ such that the variable $V_{(k)}$ has $\left| \tilde{%
\varphi}_{k}\right\rangle $ as an eigenstate. In particular, we denote the
eigenvalues of $V_{(k)}$ by $\{v_{(k),j}\}_{j=1}^{m},$ and the associated
eigenvectors by $\{\left| \varphi _{(k),j}\right\rangle \}_{j=1}^{m},$ and
take $\left| \varphi _{(k),1}\right\rangle \equiv \left| \tilde{\varphi}%
_{k}\right\rangle .$ It then follows that if the first measurement has
outcome $k,$ so that $\left| \tilde{\varphi}_{k}\right\rangle $ is prepared,
and if $V^{\prime }=V_{(k)},$ that is, the second apparatus measures $%
V_{(k)},$ then with probability $1,$ the second measurement has the outcome $%
1.$ Hence the faithfulness criterion is applicable in this case, and implies
that $V_{(k)}$ is determinate with value $v_{(k),1}$ at time $t,$
immediately prior to the second measurement.

We now introduce a critical assumption about the sequence of measurements:
the nature of the second measurement is taken to depend on the outcome of
the first. In particular, we imagine a set-up where if the outcome of the
first measurement is $k,$ then the second apparatus measures the variable $%
V_{(k)}$; we imagine that this is done mechanically by the measurement
apparatus, without the intervention of a physicist. In this case, the
faithfulness criterion is applicable for all possible outcomes of the first
measurement.

We now show that for the faithfulness criterion to be satisfied for such a
sequence of measurements, the preferred projectors must be non-orthogonal.
Since there is a non-zero probability for the first measurement to have the
outcome $k$ for every $k,$ it follows from the faithfulness criterion that
there is a non-zero probability for the system to possess the property $%
[V_{(k)}]=v_{(k),1}$ immediately prior to the second measurement, for every $%
k.$ If a property has non-zero probability of being possessed, it is a
possible property. Since the projector $P_{\tilde{\varphi}_{k}}$ associated
with the property $[V_{(k)}]=v_{(k),1}$ is one-dimensional, it has no
non-null proper subspaces, and thus by definition it is the preferred
projector for the $k$th property ascription. Thus, the set of preferred
projectors is $\{P_{\tilde{\varphi}_{k}}\}_{k=1}^{m}.$ Finally, since the
set of vectors $\{\left| \tilde{\varphi}_{k}\right\rangle \}_{k=1}^{m}$ is
by hypothesis non-orthogonal, the set of preferred projectors must also be
non-orthogonal.

For clarity, we briefly repeat this argument in the context of a simple
example, illustrated in Fig. 1. Suppose the variables being measured
correspond to the components along different spatial axes of the spin
operator, ${\bf S,}$ for a spin 1/2 particle. Denote the component along
axis ${\bf \hat{n}}$ by ${\bf S\cdot \hat{n},}$ and the eigenstate
associated with eigenvalue $\pm \hbar /2$ by $\left| \pm {\bf \hat{n}}%
\right\rangle .$ Suppose the first measurement is along ${\bf \hat{z},}$ so
that $V={\bf S\cdot \hat{z},}$ $\left| \varphi _{1}\right\rangle =\left| +%
{\bf \hat{z}}\right\rangle ,$ and $\left| \varphi _{2}\right\rangle =\left| -%
{\bf \hat{z}}\right\rangle .$ Suppose moreover that the state $\left| +{\bf 
\hat{x}}\right\rangle $ is prepared if the outcome of the first measurement
is $-{\bf \hat{z}},$ while no disturbance occurs if the outcome is $+{\bf 
\hat{z}},$ so that $\left| \tilde{\varphi}_{1}\right\rangle =\left| +{\bf 
\hat{z}}\right\rangle ,$ and $\left| \tilde{\varphi}_{2}\right\rangle
=\left| +{\bf \hat{x}}\right\rangle .$ Now suppose that the manner in which
the nature of the second measurement depends on the outcome of the first is
the following: if the first measurement has outcome $+{\bf \hat{z}},$ then
the second measurement is of ${\bf S\cdot \hat{z}}$, while if it has outcome 
$-{\bf \hat{z}}$, then the second measurement is of ${\bf S\cdot \hat{x}.}$
Thus, $V_{(1)}={\bf S\cdot \hat{z}}$ and $V_{(2)}={\bf S\cdot \hat{x}.}$
Assume the initial state of the spin is $c_{1}\left| +{\bf \hat{z}}%
\right\rangle +c_{2}\left| -{\bf \hat{z}}\right\rangle ,$ where $\left|
c_{1}\right| ^{2}+\left| c_{2}\right| ^{2}=1$ and $c_{1},c_{2}\ne 0.$

Using the generalized Born rule, it is straightforward to verify that the
result of the second measurement is predictable with probability 1 given the
result of the first measurement. It follows from the faithfulness criterion
that if the outcome of the first measurement is $+{\bf \hat{z}}$, then ${\bf %
S\cdot \hat{z}}$ is subsequently determinate with value $+\hbar /2$, and if
the outcome of the first measurement is $-{\bf \hat{z}}$, then ${\bf S\cdot 
\hat{x}}$ is subsequently determinate with value $+\hbar /2$. Since both of
these options occur with non-zero probability, the properties $[{\bf S\cdot 
\hat{z}]=+\hbar /}2$ and $[{\bf S\cdot \hat{x}]=+\hbar /}2$ are both
possible. It follows that the preferred projectors are $P_{+{\bf \hat{z}}}$
and $P_{+{\bf \hat{x}}},$ which are non-orthogonal.

\subsection{Motivation}

We now consider the reasons for adopting the faithfulness criterion. \strut
As is argued by Redhead\cite{Redhead}, in seeking a realist interpretation
of quantum mechanics one is seeking an {\em explanation} of the successes of
the operational version of the theory. One way to secure an explanation of a
measurement outcome is to demand that the properties of the systems involved 
{\em ensure} this outcome. The faithfulness criterion goes beyond this
however, in that it specifies the {\em form} that such an explanation must
take. Specifically, it is assumed that the reason a variable $V$ is found to
have value $v$ in a measurement that is predictable with probability 1 is
because immediately prior to the measurement $V$ is determinate and has
value $v$. Although this is perhaps the simplest form the explanation could
take, it is not the only form, as is evidenced by Bohm's theory\cite{Bohm}
and Bell's be-able interpretation\cite{Bell}, where only the outcomes of
measurements of certain variables (position in Bohm's case and lattice
fermion number in Bell's case), are taken to reveal pre-existing values of
these variables.

So we see that the faithfulness criterion is not a necessary feature of a
realist interpretation. Nonetheless, there have been many attempts to ensure
that the outcomes of perfectly predictable measurements {\em do} reveal
pre-existing values of these variables.

{\em \strut }This tradition dates back to von Neumann, who assumed that the
determinate variables of a system and their values are fixed by the density
operator for the system, $\rho (t),$\ by what we shall call the {\em %
orthodox rule}, namely, 
\begin{eqnarray}
Ont(t) &=&\{V|V\rho (t)\propto \rho (t)\}  \nonumber \\
\lbrack V]_{t} &=&Tr(V\rho (t)),  \label{orthodox rule}
\end{eqnarray}
where $Ont(t)$\ indicates the ontology at time $t,$\ and $[V]_{t}$\
indicates the value of $V$\ at time $t$ (this is simply the rule adopted by
the `orthodox' realist interpretations discussed in the introduction). \
Given that after a non-disturbing measurement of the variable $V$\ with
outcome $v,$\ one can predict, with probability $1,$\ that the outcome of an
immediately consecutive ideal measurement of $V$\ will also be $v$, the
faithfulness criterion demands that $V$\ be determinate with value $v$\
prior to the second measurement. However, given the orthodox rule, this can
only occur if the density operator after the first measurement is a
projector onto an eigenstate of $V$\ associated with eigenvalue $v.$\ This
must be the case regardless of the density operator prior to the first
measurement. Thus, in order to satisfy the faithfulness criterion, von
Neumann assumed that upon measurement the state vector undergoes a
non-unitary evolution (the so-called `collapse') to the eigenvector
associated with the outcome of the measurement. As a realist interpretation
of quantum mechanics this proposal is at best incomplete since it fails to
specify, in terms of the primitives of the theory, the conditions under
which a collapse occurs.

Many modal interpretretations also attempt to satisfy the faithfulness
criterion, but unlike von Neumann, they abandon the orthodox rule rather
than assuming collapse. For instance, it has occurred to many authors,
including Kochen\cite{Kochen}, Healey\cite{Healey} and Dieks\cite{Dieks
first}, that by assigning determinate status to the projectors in the
spectral resolution of the density operator one can satisfy the faithfulness
criterion for ideal measurements. Modern versions of this approach include
the proposals of Vermaas and Dieks\cite{Vermaas and Dieks}, and
Bacciagaluppi and Dickson\cite{BacciagaluppiDickson}. However, it was noted
by Bacciagaluppi and Hemmo\cite{Bacciagaluppi and Hemmo -state prep} that
the Vermaas and Dieks proposal failed to satisfy the faithfulness criterion
for certain {\em non-ideal }measurements, specifically, disturbing
measurements. The same argument can be applied against the Bacciagaluppi and
Dickson proposal.

These results do not rule out the possibility that some new proposal
involving a different, but still orthogonal, choice of preferred projectors
might satisfy the faithfulness criterion for non-ideal measurements.
However, by considering an experiment wherein the nature of the second
measurement depends on the outcome of the first, we have shown that the
faithfulness criterion fails to be satisfied for {\em any} modal
interpretation that adopts orthogonal preferred projectors.

Thus any modal proposal seeking to satisfy the faithfulness criterion must
allow for non-orthogonal preferred projectors. However, a satisfactory
proposal must provide an unambiguous rule for identifying the set of
preferred projectors for every system at every time, and it remains to be
seen whether there exists any rule that consistently satisfies faithfulness.
This rule must also satisfy other constraints, such as predicting properties
for macroscopic systems that are in accord with our everyday perceptions of
them. In particular, it must yield a solution to the measurement problem. It
may be that the preferred set {\em cannot} be chosen to satisfy faithfulness
for all measurements while also satisfying these other constraints. If this
were true, it would certainly remove some of the motivation for pursuing a
modal interpretation in the tradition of the authors specified above. We are
not able to rule out this possibility here. Nonetheless, the range of
measurements for which faithfulness is satisfied can at least be expanded if
one assumes non-orthogonal preferred projectors, as we demonstrate in
section 5 by a specific proposal,

\subsection{Consequences for the ontology}

In a modal interpretation, the property ascription to a system at a given
time can be one of several possibilities. A question which we now address is
whether or not these possibilities should differ with respect to the
ontology they ascribe. Most previous modal interpreters have assumed that
they should not. In such interpretations, the possible property ascriptions
differ only with respect to the value ascription to a single common
ontology. However, as we now prove, such an approach is unable to accomodate
non-orthogonal preferred projectors.

\begin{description}
\item[Theorem 1]  It is not possible for there to be, at a given time,
several possible mutually exclusive property ascriptions which (1) satisfy
the algebraic constraints, (2) do not differ with respect to ontology, and
(3) are associated with preferred projectors that are non-orthogonal.
\end{description}

{\bf Proof. }The proof is by contradiction. Suppose the ontology and
preferred projector for the $k$th property ascription (in the set of
possible property ascriptions at a given time) are denoted respectively by $%
Ont_{k}$ and $P_{k}.$ Since by hypothesis the possible property ascriptions
do not differ with respect to ontology, there exists a single set of
determinate variables, denoted by $Ont,$ such that $\forall k:Ont_{k}=Ont$.
Since $\forall k:P_{k}\in Ont_{k}$, it follows that $\{P_{k}\}_{k=1}^{m}\in
Ont.$ In other words, if there is only a single possible ontology at a given
time, the preferred projectors for all the different property ascriptions
must simultaneously be part of this ontology. Since one of the possible
property ascriptions must actually obtain, one of these projectors must
receive the value $1.$ Moreover, since by hypothesis the preferred
projectors are non-orthogonal, it follows that the common ontology includes
several non-orthogonal projectors, one of which receives the value 1.
However, this is in contradiction with the algebraic constraints, as we now
demonstrate.

Suppose that $P_{{\cal S}}$ and $P_{{\cal S}^{\prime }}$ are two
non-orthogonal preferred projectors, and that $[P_{{\cal S}}]=1.$ Since the
two property ascriptions associated with these are by assumption mutually
exclusive, the intersection of ${\cal S}$ and ${\cal S}^{\prime }$ is the
null space. Moreover, by the definition of a preferred projector, no
non-null proper subspace of ${\cal S}$ or ${\cal S}^{\prime }$ can be
determinate. By closure and the fact that $P_{{\cal S}^{\prime }}$ is
determinate, $P_{({\cal S}^{\prime })^{\perp }}$ is also determinate. By
closure and the fact that $P_{{\cal S}}$ and $P_{({\cal S}^{\prime })^{\perp
}}$ are determinate, the projector onto the intersection of $({\cal S}%
^{\prime })^{\perp }$ and ${\cal S}$ must also be determinate. Since no
non-null proper subspaces of ${\cal S}$ can be determinate, the intersection
of $({\cal S}^{\prime })^{\perp }$ and ${\cal S}$ cannot be a non-null
proper subspace of ${\cal S}.$ Moreover, this intersection cannot be ${\cal S%
}$ itself, since then ${\cal S}$ and ${\cal S}^{\prime }$ would be
orthogonal, contradicting our initial assumption. Thus, the intersection of $%
({\cal S}^{\prime })^{\perp }$ and ${\cal S}$ must be the null space. It
then follows from the functional relation constraint that $[P_{({\cal S}%
^{\prime })^{\perp }}][P_{{\cal S}}]=0,$ and since $[P_{{\cal S}}]=1,$ this
implies that $[P_{({\cal S}^{\prime })^{\perp }}]=0.$ It also follows from
the functional relation constraint that $[P_{({\cal S}^{\prime })^{\perp
}}]=1-[P_{{\cal S}^{\prime }}]$, so that $[P_{{\cal S}^{\prime }}]=1.$ Thus,
both $P_{{\cal S}}$ and $P_{{\cal S}^{\prime }}$ receive the value $1$. But,
this is in contradiction with $[P_{{\cal S}}][P_{{\cal S}^{\prime }}]=0$
which follows from the fact that the intersection of ${\cal S}$ and ${\cal S}%
^{\prime }$ is the null space$.$ {\bf QED.}

\section{An Interpretive Framework Incorporating Non-orthogonal Preferred
Projectors}

\subsection{Preliminaries}

In the previous section it was established that in order to consider a modal
interpretation with non-orthogonal preferred projectors, the different
possible property ascriptions to a system must differ with respect to the
ontology they ascribe. The precise form of the ontology associated with a
particular property ascription has not yet been specified. It turns out that
this form is fixed if an additional constraint on the property ascription is
adopted, namely,

\begin{description}
\item[Weakening Condition]  \ If $P_{{\cal S}}\in Ont$ and $[P_{{\cal S}}]=1$
then for all $P_{{\cal R}}$ such that $P_{{\cal R}}\ge P_{{\cal S}},$ $P_{%
{\cal R}}\in Ont$ and $[P_{{\cal R}}]=1.$
\end{description}

{\em \ }This condition was introduced by Healey\cite{Healey} and was
resurrected recently by Vermaas\cite{Vermaas expanding}. It is called
`weakening' since in Healey's terminology $P_{{\cal R}}$ is said to be
weaker than $P_{{\cal S}}$ if $P_{{\cal R}}\ge P_{{\cal S}}$. It is
motivated by the same sorts of considerations that lead one to adopt the
closure constraint and the functional relation constraint; it is an attempt
to preserve the logical structure of classical mechanics. In the language of
properties, the weakening condition states that if property $s$ is
well-defined and holds for the system, then any property implied by $s,$
namely any property of the form `$s$ or $s^{\prime }$' should also be
well-defined and hold for the system. It should be noted that the weakening
condition is unlike previous constraints, insofar as the nature of the
ontology is made to depend on features of the value ascription.

We now demonstrate the form of property ascription that results from
adopting the weakening condition.

\begin{description}
\item[Theorem 2]  \ The algebraic constraints and the weakening condition
imply that the set of determinate projectors and the value ascription to
these must respectively have the forms 
\begin{eqnarray}
\{P_{{\cal R}}|P_{{\cal R}} &\ge &P_{{\cal S}}\text{ or }P_{{\cal R}}\perp
P_{{\cal S}}\}  \label{determinate projectors} \\
\lbrack P_{{\cal R}}] &=&\left\{ 
\begin{array}{c}
1\text{ if }P_{{\cal R}}\ge P_{{\cal S}} \\ 
0\text{ if }P_{{\cal R}}\perp P_{{\cal S}}
\end{array}
\right. ,  \label{value ascription to determinate projectors}
\end{eqnarray}
where $P_{{\cal S}}$ is the preferred projector for the property ascription.
\end{description}

{\bf Proof.} Recall that the preferred projector $P_{{\cal S}}$ is the
unique projector in the property ascription satisfying Eq. (\ref{defintion
of pref proj}), so that $[P_{{\cal S}}]=1$ and there is no subspace ${\cal T}
$ such that $P_{{\cal T}}<P_{{\cal S}}$ and $[P_{{\cal T}}]=1.$ By the
weakening condition, $[P_{{\cal S}}]=1$ implies that the set of projectors $%
\{P_{{\cal R}}|P_{{\cal R}}\ge P_{{\cal S}}\}$ is determinate. Moreover, for
any projector $P_{{\cal U}}$ orthogonal to $P_{{\cal S}},$ $P_{{\cal U}}+P_{%
{\cal S}}$ is determinate, since $(P_{{\cal S}}+P_{{\cal U}})\in \{P_{{\cal R%
}}|P_{{\cal R}}\ge P_{{\cal S}}\}.$ It then follows from the constraint of
closure that $(I-P_{{\cal S}})(P_{{\cal S}}+P_{{\cal U}})=P_{{\cal U}}$ is
determinate. Thus, all projectors orthogonal to $P_{{\cal S}},$ namely the
set $\{P_{{\cal U}}|P_{{\cal U}}\perp P_{{\cal S}}\},$ are also determinate.
In summary, all the projectors in{\em \ }the set $\{P_{{\cal R}}|P_{{\cal R}%
}\ge P_{{\cal S}}$ or $P_{{\cal R}}\perp P_{{\cal S}}\}$ must be
determinate. We now show that the projectors in this set are the {\em only }%
projectors that are determinate.

Suppose the contrary, namely that there exists a determinate projector $P_{%
{\cal V}}$ such that $P_{{\cal V}}\ngeq P_{{\cal S}}$ and $P_{{\cal V}}\not%
{\perp}P_{{\cal S}}.$ If $P_{{\cal V}}\ngeq P_{{\cal S}}$ then the
intersection of ${\cal V}$ and ${\cal S}$ is not equal to ${\cal S}$, and
must therefore be a proper subspace of ${\cal S}.$ It then follows from the
functional relation constraint and the assumption that all proper subspaces
of ${\cal S}$ receive the value $0$ that $[P_{{\cal V}}][P_{{\cal S}}]=0.$
Since $[P_{{\cal S}}]=1,$ we conclude that $[P_{{\cal V}}]=0.$ Moreover,
since $P_{{\cal V}}\not{\perp}P_{{\cal S}}$ is equivalent to $(I-P_{{\cal V}%
})\ngeq P_{{\cal S}},$ it follows by the same argument that $[I-P_{{\cal V}%
}]=0.$ But $[I-P_{{\cal V}}]=0$ implies $[P_{{\cal V}}]=1,$ thereby yielding
a contradiction.

Finally, we demonstrate that the value ascription must be of the form of (%
\ref{value ascription to determinate projectors}). Given that $[P_{{\cal S}%
}]=1,$ it follows trivially from the weakening condition that $[P_{{\cal R}%
}]=1$ if $P_{{\cal R}}\ge P_{{\cal S}}.$ Moreover, $[P_{{\cal R}}]=0$ if $P_{%
{\cal R}}\perp P_{{\cal S}}$ since otherwise the intersection of ${\cal R}$
and ${\cal S},$ which is the null space, would receive the value $1.$ {\bf %
QED.}

Theorem 2 identifies the set of idempotent variables that are determinate
given the weakening condition and the algebraic constraints. The set of
non-idempotent variables that are determinate then follows from the spectral
constraint. Specifically, we have

\begin{description}
\item[Corollary]  If the set of determinate projectors and their values are
given by Eqs.(\ref{determinate projectors}) and (\ref{value ascription to
determinate projectors}), then the spectral constraint implies that the
ontology and its value ascription are given by 
\begin{eqnarray}
Ont &=&\{V|VP_{{\cal S}}\propto P_{{\cal S}}\}  \label{ontology given root}
\\
\lbrack V] &=&Tr\left( VP_{{\cal S}}\right) ,
\label{value ascription to ontology given root}
\end{eqnarray}
where $P_{{\cal S}}$ is the preferred projector for the property ascription.
\end{description}

{\bf Proof.} The spectral constraint states that a non-idempotent variable $%
V $ is determinate if and only if all the elements of its spectral
resolution are determinate. Thus every variable $V$ that is determinate has
a spectral resolution $V=\sum_{k}\lambda _{k}P_{{\cal R}_{k}},$ where $%
\forall k:$ $(P_{{\cal R}_{k}}\ge P_{{\cal S}}$ or $P_{{\cal R}_{k}}\perp P_{%
{\cal S}}).$ But the latter condition is equivalent to $\forall k:$ $P_{%
{\cal R}_{k}}P_{{\cal S}}\propto P_{{\cal S}},$ from which it follows that $%
VP_{{\cal S}}\propto P_{{\cal S}}.$ The spectral constratint also states
that $[V]=\sum_{k}\lambda _{k}[P_{{\cal R}_{k}}]$. Since the $P_{{\cal R}%
_{k}}$ are orthogonal, only one can satisfy $P_{{\cal R}_{k}}\ge P_{{\cal S}%
},$ and thereby receive the value $1$ by Eq.(\ref{value ascription to
determinate projectors}). Labelling this projector by $k^{\prime },$ we have 
$[V]=\lambda _{k^{\prime }},$ and $VP_{{\cal S}}=\lambda _{k^{\prime }}P_{%
{\cal S}},$ from which Eq.(\ref{value ascription to ontology given root})
follows. {\bf QED.}

The corollary to theorem 2 states that a variable is determinate if it has
the subspace associated with the preferred projector as an eigenspace, and
the value of this variable is the associated eigenvalue. This has the form
of the orthodox rule, defined in Eq. (\ref{orthodox rule}), but where the
role of the density operator is played by the preferred projector.

Note that for systems of dimensionality 3 or greater, theorem 2 implies that
the ontologies associated with mutually exclusive property ascriptions are
necessarily distinct. This holds true for such systems even if the preferred
projectors for the property ascriptions are orthogonal\footnote{%
This is not true for a 2-dimensional Hilbert space, since two distinct
property ascriptions can be associated with the same ontology. This occurs
when the preferred projectors for these property ascriptions are orthogonal.}%
. In this sense, the weakening condition provides another reason,
independent of the one provided in section 3.3, for allowing the possible
property ascriptions to differ in ontology. Such an argument was in fact
made by Vermaas in the context of the Vermaas and Dieks version of the modal
interpretation \cite{Vermaas expanding}.

Thus far, we have focused upon the property ascriptions for individual
systems, and nothing has been said concerning the relationship between the
property ascriptions to composite systems and the subsystems of which they
are formed. {\em \strut }Clifton\cite{Cliftonproperties} has argued for the
following constraint on this relationship, which we call the {\em %
reductionist rule}:

\begin{equation}
V^{A}\in Ont^{A}\text{ if and only if }V^{A}\otimes I^{B}\in Ont^{AB},\text{
and }[V^{A}]=[V^{A}\otimes I^{B}],  \label{reductionist rule}
\end{equation}
where $Ont^{A}$ is the ontology of system $A,$ and $AB$ is the composite of
systems $A$ and $B$ (i.e. $A(B)$ denotes the system associated with Hilbert
space ${\cal H}^{A}({\cal H}^{B}),$ and $AB$ denotes the system associated
with ${\cal H}^{A}\otimes {\cal H}^{B}).$ Denying this constraint leads to
what Clifton has called {\it ontological perpectivalism}, the view that what
exists depends on the level of compositeness of the description. For
instance, to deny the `only if' half of the rule amounts to claiming that it
is possible for part $A$ of a composite to have the property $s$, while the
composite itself does not have the property that part $A$ has property $s$.
Clifton has characterized such a position as `metaphysically untenable'.

If one adopts the reductionist rule, the property ascriptions for all
subsystems are uniquely fixed by the property ascription for the composite.
However, we also wish to assume that the property ascriptions for {\em every}
system satisfy the algebraic constraints and the weakening condition. It has
yet to be demonstrated that these constraints are consistent with the
reductionist rule. In fact they are. Specifically, if the property
ascription for the composite has the form given in Eqs. (\ref{ontology given
root}) and (\ref{value ascription to ontology given root}), with a preferred
projector denoted by $P_{S}^{AB},$ then the property ascription for
subsystem $A$ also has the form given in Eqs. (\ref{ontology given root})
and (\ref{value ascription to ontology given root}), where the preferred
projector, denoted by $P_{{\cal S}}^{A},$ is the unique projector satisfying 
\begin{equation}
P_{{\cal S}}^{AB}\le P_{{\cal S}}^{A}\otimes I^{B}\text{ and there is no
subspace }{\cal T}^{A}\text{ such that }P_{{\cal S}}^{AB}\le P_{{\cal T}%
}^{A}\otimes I^{B}<P_{{\cal S}}^{A}\otimes I^{B}.
\label{pref proj for subsystem}
\end{equation}
(In other words, $P_{{\cal S}}^{A}$ is the `smallest' projector satisfying $%
P_{{\cal S}}^{A}\otimes I^{B}\ge P_{{\cal S}}^{AB})$ This can be shown to be
a limiting case of a result by Dickson and Clifton \cite{Dickson and Clifton}%
, however for clarity we prove it directly. It suffices to demonstrate the
following equivalences: 
\[
\{P_{{\cal R}}^{A}|P_{{\cal R}}^{A}\otimes I^{B}\ge P_{{\cal S}}^{AB}\}=\{P_{%
{\cal R}}^{A}|P_{{\cal R}}^{A}\ge P_{{\cal S}}^{A}\} 
\]
and 
\[
\{P_{{\cal R}}^{A}|\text{ }P_{{\cal R}}^{A}\otimes I^{B}\perp P_{{\cal S}%
}^{AB}\}=\{P_{{\cal R}}^{A}|P_{{\cal R}}^{A}\perp P_{{\cal S}}^{A}\}, 
\]
where $P_{{\cal S}}^{A}$ and $P_{{\cal S}}^{AB}$ are related as above. We
first demonstrate that the right hand sides imply the left. $P_{{\cal R}%
}^{A}\ge P_{{\cal S}}^{A}$ trivially implies $P_{{\cal R}}^{A}\otimes
I^{B}\ge P_{{\cal S}}^{A}\otimes I^{B},$ and since by definition, $P_{{\cal S%
}}^{A}\otimes I^{B}\ge P_{{\cal S}}^{AB},$ it follows that $P_{{\cal R}%
}^{A}\otimes I^{B}\ge P_{{\cal S}}^{AB}.$ Similarly, $P_{{\cal R}}^{A}\perp
P_{{\cal S}}^{A}$ trivially implies $P_{{\cal R}}^{A}\otimes I^{B}\perp P_{%
{\cal S}}^{A}\otimes I^{B},$ and together with $P_{{\cal S}}^{A}\otimes
I^{B}\ge P_{{\cal S}}^{AB},$ this implies that $P_{{\cal R}}^{A}\otimes
I^{B}\perp P_{{\cal S}}^{AB}.$ To show that the left hand sides imply the
right, we make use of the fact that $P_{{\cal S}}^{A}\otimes I^{B}$ is the
`smallest' projector satisfying $P_{{\cal S}}^{A}\otimes I^{B}\ge P_{{\cal S}%
}^{AB}.$ This implies that any projector $P_{{\cal R}}^{A}$ satisfying $P_{%
{\cal R}}^{A}\otimes I^{B}\ge P_{{\cal S}}^{AB}$ must also satisfy $P_{{\cal %
R}}^{A}\otimes I^{B}\ge P_{{\cal S}}^{A}\otimes I^{B},$ and hence $P_{{\cal R%
}}^{A}\ge P_{{\cal S}}^{A}.$ In addition, any projector $P_{{\cal R}}^{A}$
satisfying $P_{{\cal R}}^{A}\otimes I^{B}\perp P_{{\cal S}}^{AB}$
(equivalently $(I^{A}-P_{{\cal R}}^{A})\otimes I^{B}\ge P_{{\cal S}}^{AB})$
must also satisfy $(I^{A}-P_{{\cal R}}^{A})\otimes I^{B}\ge P_{{\cal S}%
}^{A}\otimes I^{B},$ which implies $P_{{\cal R}}^{A}\perp P_{{\cal S}}^{A}.$
This concludes the proof.

Clearly, the preferred projectors for a subsystem can be non-orthogonal if
the preferred projectors for the composite are non-orthogonal. What is
perhaps more surprising is that the preferred projectors for a subsystem can
be non-orthogonal even if the preferred projectors for the composite are
not! For example, suppose the preferred projectors for the composite are two
orthogonal projectors, $P_{1}^{AB}=P_{1}^{A}\otimes P_{1}^{B}$ and $%
P_{2}^{AB}=P_{2}^{A}\otimes P_{2}^{B},$ where $P_{1}^{B}$ and $P_{2}^{B}$
are orthogonal projectors, but $P_{1}^{A}$ and $P_{2}^{A}$ are not. It then
follows from Eq. (\ref{pref proj for subsystem}) that the preferred
projectors for $A$ are simply $P_{1}^{A}$ and $P_{2}^{A},$ which are
non-orthogonal.

In the next subsection, we will introduce a framework for interpretation
wherein the possible property ascriptions for the universe are defined
first, in accordance with the algebraic constraints and the weakening
condition, and the possible property ascriptions for all subsystems are then
inferred using the reductionist rule. The preferred projectors for the
universe will be assumed to be orthogonal, but as shown above, this is
consistent with the preferred projectors for a subsystem being
non-orthogonal and hence does not rule out the possibility of satisfying the
faithfulness criterion. A more general approach would be to assume a
non-orthogonal preferred set for the universe as well. However, the
faithfulness criterion does not necessitate this assumption, and indeed, as
we will demonstrate in section 5, one can satisfy this criterion for a wide
variety of measurements without it. The case of an orthogonal preferred set
for the universe is in any event a natural place to begin such an
investigation.

The framework that emerges is similar to the one proposed by Bub and Clifton%
\cite{Bub and Clifton}. The most significant difference is in the form of
the property ascription, since the latter do not assume the weakening
condition. Another difference is in the dynamics of the property ascription.
Bub and Clifton defined a dynamics following Vink\cite{Vink} and Bell\cite
{Bell}. This approach was subsequently generalized in two respects by
Bacciagaluppi and Dickson\cite{BacciagaluppiDickson}, and Dickson\cite
{Dicksonplurality}. First, the preferred projectors were allowed to be
time-dependent, and secondly it was shown that there is a plurality of
possible dynamics consistent with the quantum statistics. We follow the
latter, generalized approach.

Since many of the ingredients of the framework derive from a number of
sources, and since we introduce some novel terminology, we have written the
rest of this section in such a way that it constitutes a self-contained
description of the framework.

\subsection{Details of the framework}

It is assumed that the universe is associated with a Hilbert space ${\cal H}$
and a vector $\left| \psi (t)\right\rangle \in {\cal H}$ that evolves
deterministically over time in accordance with the Schr\"{o}dinger equation, 
\begin{equation}
\frac{d}{dt}\left| \psi (t)\right\rangle =-iH\left| \psi (t)\right\rangle ,
\label{Schrodinger}
\end{equation}
where $H$ is the total Hamiltonian, and where the units are chosen such that 
$\hbar =1$. Since, as will be demonstrated shortly, the role of the vector $%
\left| \psi (t)\right\rangle $ in the framework is to determine the
probabilities of various different property ascriptions as well as their
dynamics, it will be dubbed the{\em \ dynamical state vector.}

Define a decomposition $D$ of a vector $\left| \psi \right\rangle $ as a set 
$\{(c_{k},\left| \phi _{k}\right\rangle )\}_{k=1}^{m}$ of non-zero
coefficients $c_{k}$ and orthonormal vectors $\left| \phi _{k}\right\rangle $
such that $\left| \psi \right\rangle =\sum_{k=1}^{m}c_{k}\left| \phi
_{k}\right\rangle .$ It is assumed that every interpretation within the
framework selects a {\em preferred decomposition} of the dynamical state $%
\left| \psi (t)\right\rangle $ at every time $t.$ The projectors onto the
elements of the preferred decomposition constitute the preferred projectors
for the possible property ascriptions to the universe.

We also introduce a new `state vector' that we denote by $\left| \Phi
(t)\right\rangle .$ It can be any one of the vector elements of the
preferred decomposition. The projector onto this vector is the preferred
projector for the property ascription to the universe that obtains at time $%
t ${\em . }Assuming the algebraic constraints and the weakening condition,
it follows from Theorem 2 that the property ascription for the universe has
the form of Eqs. (\ref{ontology given root}) and (\ref{value ascription to
ontology given root}), which may be rewritten in terms of $\left| \Phi
(t)\right\rangle $ as 
\begin{eqnarray*}
Ont(t) &=&\{V|V\left| \Phi (t)\right\rangle \propto \left| \Phi
(t)\right\rangle \}, \\
\lbrack V]_{t} &=&\left\langle \Phi (t)\right| V\left| \Phi (t)\right\rangle
.
\end{eqnarray*}
The property ascription to any subsystem of the universe is then fixed by
the reductionist rule, defined in Eq. (\ref{reductionist rule}). Since $%
\left| \Phi (t)\right\rangle $ determines the property ascription to every
system, we call it {\em property state vector.}

Next, we introduce a restriction on the manner in which the elements of the
preferred decomposition can evolve over time. Suppose the set of vectors $%
\{\left| \phi _{k}(t)\right\rangle \}_{k=1}^{d}$ at every time $t$ is a
complete orthogonal basis for ${\cal H}$ that includes as a subset the
vector elements of the preferred decomposition. We require that there is an
indexing of the basis vectors such that every vector with a given index is
an analytic function of time. We call this the {\em constraint of
analyticity. }It can be satisfied by requiring that the time-dependent
vectors in the set $\{\left| \phi _{k}(t)\right\rangle \}_{k=1}^{d}$ each
define a path through Hilbert space obeying the equation 
\begin{equation}
\frac{d}{dt}\left| \phi _{k}(t)\right\rangle =-i\tilde{H}(t)\left| \phi
_{k}(t)\right\rangle ,  \label{analyticity}
\end{equation}
for some Hermitian operator $\tilde{H}(t).$ It is convenient to refer to
these vectors, considered as functions of time, as the {\em preferred paths}.

It is assumed that the property state vector evolves according to a
Markovian stochastic dynamics that permits hopping among the preferred
paths. We require that at every time $t$ the probability $p_{k}(t)$ that the
property state vector lies on the $k$th preferred path is given by 
\begin{equation}
p_{k}(t)=\left| \left\langle \phi _{k}(t)|\psi (t)\right\rangle \right| ^{2}.
\label{born}
\end{equation}
The latter requirement is called the {\em Born rule constraint}. Although
the basis $\{\left| \phi _{k}(t)\right\rangle \}_{k=1}^{d}$ that is defined
at time $t$ by the preferred paths may include elements that are {\em not}
part of the preferred decomposition, these elements have no overlap with $%
\left| \psi (t)\right\rangle ,$ so that the probability associated with them
is zero. It follows therefore that the property state vector always
corresponds to one of the vector elements of the preferred decomposition.
There are many dynamics that satisfy the Born rule constraint; these will be
considered in the next subsection.

We refer to this entire interpretive structure as a `framework' for modal
interpretations, since there are a plurality of possible interpretations
that have this form. Specifically, there is a different interpretation for
every choice of rule for determining the preferred decomposition and every
choice of dynamics that satisfies the Born rule constraint.

\subsection{The general form of the dynamics}

We now recall the general form of a Markovian stochastic dynamics that
satisfies the Born rule constraint\cite{BacciagaluppiDickson}. This
constraint, articulated in Eq.(\ref{born}), can be recast as constraints
upon the initial conditions\cite{distribution postulate} and the dynamics: 
\[
p_{k}(0)=\left| \left\langle \phi _{k}(0)|\psi (0)\right\rangle \right|
^{2}, 
\]
and 
\begin{equation}
\frac{d}{dt}p_{k}(t)=\frac{d}{dt}\left| \left\langle \phi _{k}(t)|\psi
(t)\right\rangle \right| ^{2}.  \label{dp/dt}
\end{equation}
Using Eqs.(\ref{Schrodinger}) and (\ref{analyticity}), the latter becomes 
\begin{equation}
\frac{d}{dt}p_{k}(t)=2%
\mathop{\rm Im}%
\left[ \left\langle \psi (t)|\phi _{k}(t)\right\rangle \left\langle \phi
_{k}(t)|H-\tilde{H}(t)|\psi (t)\right\rangle \right] .  \label{stuff}
\end{equation}

Since we assume Markovian dynamics, it is sufficient to specify the
probability $T_{kj}(t)dt$ of a transition from path $j$ to path $k$ during
the infinitesimal interval between $t$ and $t+dt,$ for all $j$ and $k.$ The
evolution of a probability distribution $p_{k}(t)$ over the paths is then
given by the master equation 
\[
\frac{d}{dt}p_{k}(t)=\sum_{j}\left[
T_{kj}(t)p_{j}(t)-T_{jk}(t)p_{k}(t)\right] . 
\]
In what follows, we consider the problem of finding a set of functions $%
T_{kj}$\strut $(t)$ that satisfy the master equation given $p_{k}(t).$
Following Bell\cite{Bell}, it is useful to define a new set of functions,
namely a set of probability currents, $J_{kj}(t),$ as follows: 
\begin{equation}
J_{kj}(t)=T_{kj}(t)p_{j}(t)-T_{jk}(t)p_{k}(t).  \label{current}
\end{equation}
The current $J_{kj}(t)$ describes the net flow of probability from path $j$
to $k$ at time $t.$ This definition implies that the current is
antisymmetric with respect to an interchange of its indices 
\begin{mathletters}
\begin{equation}
J_{kj}(t)=-J_{jk}(t).  \label{antisym}
\end{equation}
In terms of these currents, the master equation becomes a continuity
equation: 
\end{mathletters}
\begin{equation}
\frac{d}{dt}p_{k}(t)=\sum_{j}J_{kj}(t).  \label{continuity}
\end{equation}
Following Bacciagaluppi and Dickson\cite{BacciagaluppiDickson}, one can
solve for the $T_{kj}(t)$ in two steps. First, one finds a set of currents $%
J_{kj}(t)$ that satisfy Eq.(\ref{antisym}) and that solve Eq.(\ref
{continuity}) with $dp_{k}(t)/dt$ given by Eq.(\ref{stuff}). Next, one finds
a set of functions $T_{kj}(t)$ that solve Eq.(\ref{current}) given a
particular solution for $J_{kj}(t).$ It turns out that there an infinite
number of sets of antisymmetric currents which solve the continuity
equation. Moreover, for a given set of currents, there are an infinite
number of solutions for the $T_{kj}(t),$ specifically, any set of functions
that satisfy 
\begin{equation}
T_{kj}\ge \max \{0,\frac{J_{kj}}{p_{j}}\},  \label{Tnm}
\end{equation}
\qquad \qquad and 
\begin{mathletters}
\begin{equation}
T_{jk}=\frac{(T_{kj}p_{j}-J_{kj})}{p_{k}}.  \label{Tmn}
\end{equation}
for every pair of indices $k>j$.

\strut So we see that there is a large number of solutions for the dynamics
which satisfy the constraints introduced. It is possible that additional
constraints, such as a requirement of quantum-classical correspondence,
might eliminate the ambiguity in the choice of dynamics, but this has yet to
be demonstrated and some authors argue that it is unlikely\cite
{Dicksonplurality}.

\section{The Minimal Entropy Proposal}

\subsection{Details of the proposal}

We begin by introducing some terminology. A {\it factorization}{\bf \ }$F$
of a Hilbert space ${\cal H}$ is defined to be a set of Hilbert spaces each
of dimensionality greater than one, the direct product of which is ${\cal H}%
, $ that is, $F$ $=\{{\cal H}^{(p)}\}_{p=1}^{n},$ such that ${\cal H}={\cal H%
}^{(1)}\otimes {\cal H}^{(2)}\cdot \cdot \cdot \otimes {\cal H}^{(n)},$ and $%
\dim ({\cal H}^{(p)})>1$. A more precise definition of this concept is
supplied by Bacciagaluppi\cite{Bacciagaluppi no-go}, but this is not
required for our purposes. A factorization containing $n$ elements is called 
$n$-{\it partite, }and the elements themselves are called {\it factor spaces.%
} A factorization $F$ is said to be a {\it coarse-graining}{\bf \ }of a
factorization $F^{\prime },$ and $F^{\prime }$ a {\it fine-graining} of $F,$
if $F^{\prime }$ can be generated from $F$ by factorizing one or more of the
elements of $F.$ Finally, a {\it product decomposition} of $\left| \psi
\right\rangle $ with respect to the factorization $F=\{{\cal H}%
^{(p)}\}_{p=1}^{n}$ is any decomposition $\{(c_{k},\bigotimes_{p=1}^{n}%
\left| \phi _{k}^{(p)}\right\rangle )\}_{k=1}^{m}$ of $\left| \psi
\right\rangle $ every element of which is a product state over $F.$

The first element of the proposal is to assume that there is a factorization
of the Hilbert space of the universe that is more physically relevant than
the others; we call it the {\em distinguished factorization} . There is a
precedent for such an assumption, specifically, in the modal interpretations
of Healey\cite{Healey}, Bacciagaluppi and Dickson\cite{BacciagaluppiDickson}%
, and Dieks\cite{Dieks pref fact}. Such interpretations have been called
`atomic', since the factor spaces of the distinguished factorization
represent the most elementary physical systems. Some restrictions on what
the distinguished factorization could be will be discussed briefly in
section 5.2.

The first constraint upon the preferred decomposition is that it be a
product decomposition with respect to the distinguished factorization. This
constraint is not sufficient to uniquely specify a decomposition. Indeed,
the number of product decompositions of any state vector with respect to a
given factorization is infinite. In order to distinguish between these, we
turn our attention towards the coefficients in the decomposition. Since
these coefficients define a probability distribution, different
decompositions can be ordered with respect to the uniformity of the
associated distributions. This uniformity can be quantified by several
`entropic' quantities. The most obvious candidate is the Shannon entropy,
defined for a probability distribution ${\bf p}=(p_{1},p_{2},...,p_{m})$ as 
\end{mathletters}
\begin{equation}
H({\bf p)=-}\sum_{k=1}^{m}p_{k}\log p_{k}.  \label{shannonentropy}
\end{equation}
Thus, we can associate with every decomposition $D=\{(c_{k},\left| \phi
_{k}\right\rangle )\}_{k=1}^{m}$ of the state vector $\left| \psi
\right\rangle $ the entropy 
\begin{equation}
S_{\left| \psi \right\rangle }(D)={\bf -}\sum_{k=1}^{m}\left| c_{k}\right|
^{2}\log \left| c_{k}\right| ^{2}.  \label{ontological entropy}
\end{equation}
We refer to this quantity as the ${\em IU}${\em \ entropy} of the state
vector $\left| \psi \right\rangle $ for the decomposition $D,$ since it has
previously been considered by Ingarden and Urbanik\cite{Ingarden and Urbanik}%
, albeit in a very different context.

It is now possible to state our choice of preferred decomposition:

\begin{quote}
Given a distinguished factorization $F$, the preferred decomposition of the
dynamical state vector $\left| \psi \right\rangle $ is the one that
minimizes the IU entropy of $\left| \psi \right\rangle $ from among all
product decompositions with respect to $F$.
\end{quote}

By choosing the product decomposition that minimizes the IU entropy, we are
choosing the interpretation where the probability distribution over the
possible property state vectors is as narrow as possible. Moreover, since
the minimum IU entropy (from among IU entropies for product decompositions)
is zero if and only if $\left| \psi \right\rangle $ is a product state, it
can be thought of as a measure of the entanglement of $\left| \psi
\right\rangle $ with respect to the distinguished factorization. The
strongest motivation for such a choice of preferred decomposition is that it
appears very promising in securing a solution to the measurement problem and
in satisfying the faithfulness criterion, as will be demonstrated in
sections 5.2 and 5.3. We do not however offer any {\em a priori }%
justification of the principle.

Implementing the proposal requires solving the minimization problem for a
given dynamical state vector and a given choice of distinguished
factorization. If the distinguished factorization is bi-partite, the
solution is given by the following theorem.

\begin{description}
\item[Theorem 3]  \ \strut Suppose $\left| \psi \right\rangle $ is any
vector in a Hilbert space with a bi-partite distinguished factorization $F_{%
\text{bi}}.$ Any decomposition of $\left| \psi \right\rangle $ that is
bi-orthogonal with respect to $F_{\text{bi}}$ minimizes the IU entropy from
among all product decompositions of $\left| \psi \right\rangle $ with
respect to $F_{\text{bi}}.$
\end{description}

The proof of this theorem is relegated to appendix A. In the case of an $n$%
-partite distinguished factorization, with $n>2,$ we have not yet found a
solution to the minimization problem for all state vectors. However, the
bi-partite result can be used to identify the preferred decomposition for 
{\it some} state vectors, as follows.

\begin{description}
\item[Theorem 4]  \ Suppose $\left| \psi \right\rangle $ is a vector in a
Hilbert space with an $n$-partite distinguished factorization, $F_{n},$
where $n>2.$ If there exists a decomposition of $\left| \psi \right\rangle $
that is a product decomposition with respect to $F_{n}$ and that is a
bi-orthogonal decomposition with respect to some bi-partite coarse-graining
of $F_{n}$, then this decomposition minimizes the IU entropy from among all
product decompositions with respect to $F_{n}$.
\end{description}

{\bf Proof.}{\em \ }Suppose $F_{\text{bi}}$ is a bi-partite coarse-graining
of $F_{n}.$ The set $S_{n}$ of decompositions of $\left| \psi \right\rangle $
that are product decompositions with respect to $F_{n}$ is a subset of the
set $S_{\text{bi}}$ that are product decompositions with respect to $F_{%
\text{bi}}.$ We can denote this by $S_{n}\subseteq S_{\text{bi}}.$ Moreover,
suppose $D_{n}^{\min }(D_{\text{bi}}^{\min })$ is the decomposition that
minimizes the IU entropy from among all the elements of $S_{n}(S_{\text{bi}%
}).$ Theorem 3 shows that for every state vector $\left| \psi \right\rangle $%
, $D_{\text{bi}}^{\min }$ is the bi-orthogonal decomposition of $\left| \psi
\right\rangle $. For certain state vectors, it may happen that $D_{\text{bi}%
}^{\min }$ lies among the elements of $S_{n}.$ Since we know that $D_{\text{%
bi}}^{\min }$ minimizes the IU entropy from among all the elements of $S_{%
\text{bi}},$ and $S_{n}\subseteq S_{\text{bi}},$ it follows that in this
case $D_{\text{bi}}^{\min }$ also minimizes the IU entropy from among all
the elements of $S_{n}.$ Thus, in this case $D_{n}^{\min }=D_{\text{bi}%
}^{\min }.$ {\bf QED.}

Theorem 4 is not a complete solution to the minimization problem because
there exist state vectors for which $D_{\text{bi}}^{\min }$ does not lie
among the elements of $S_{n}.$ Further work is required to determine the
decomposition that minimizes the IU entropy in such cases.

We note that in the proof of theorem 3, presented in appendix A, the only
relevant feature of the IU entropy is that it has the form $%
\sum_{k=1}^{m}f(\left| c_{k}\right| ^{2})$ for some concave function $f.$ It
follows that one would obtain the same results if, instead of minimizing the
IU entropy, one minimized any other entropic quantity having this form.
However, there is no guarantee that this insensitivity to the choice of
entropic quantity persists in the more general case of state vectors for
which theorem 4 does not apply.

A possible difficulty with the minimal entropy proposal as it stands has to
do with the uniqueness of the preferred decomposition. It is well known that
the bi-orthogonal decomposition of a state vector is not unique when the
eigenvalues of the reduced density operator for one of the subsystems are
degenerate. It follows from theorem 3 that if the distinguished
factorization is bi-partite, then the decomposition that minimizes the IU
entropy may not be unique, and the minimal entropy proposal may fail to
uniquely specify a preferred decomposition. For instance, this occurs if the
dynamical state vector is the EPR-Bell state for two spins $\left| \psi
\right\rangle =2^{-1/2}(\left| +{\bf a}\right\rangle \left| -{\bf a}%
\right\rangle +\left| -{\bf a}\right\rangle \left| +{\bf a}\right\rangle ).$
This difficulty persists in the case of an $n$-partite distinguished
factorization, $F_{n},$ where $n>2$, since there are dynamical state vectors
for which theorem 4 applies and the decomposition that minimizes the IU
entropy is non-unique; an example being a tensor product of EPR-Bell states.
It should be noted however that a degeneracy among the eigenvalues of the
reduced density operator for one of the factor spaces of $F_{n}$ does not 
{\em always} lead to a non-unique preferred decomposition. For instance, if
the dynamical state vector has a decomposition that is $n$-orthogonal with
respect to the factorization $F_{n},$ then it follows from theorem 4 that
this decomposition minimizes the IU entropy, and since the $n$-orthogonal
decomposition is unique for $n>2$ \cite{Peres n-orthogonal}, so is the
preferred decomposition$.$ It is an open question whether the minimization
of the IU entropy leads to a unique preferred decomposition when the
dynamical state vector is such that theorem 4 does not apply.

It is useful to distinguish two cases of non-uniqueness of the preferred
decomposition: an instantaneous non-uniqueness, occurring at an isolated
moment in time, and an extended non-uniqueness, occurring over a finite
interval of time. If the constraint of analyticity (defined in Eq.(\ref
{analyticity})) holds for the minimal entropy proposal, then the
instantaneous non-uniqueness problem can be solved easily: the preferred
paths at the moment of non-uniqueness are simply taken to be the limit of
the preferred paths at adjoining times. This is the same solution as was
proposed in the context of the atomic modal interpretation by Bacciagaluppi
and Dickson\cite{BacciagaluppiDickson}. The extended non-uniqueness problem
is not so easily solved. One possible approach to the problem is to argue
that cases wherein there is an extended non-uniqueness have negligible
probability. Since such an argument has been made for the occurrence of a
non-unique bi-orthogonal decomposition by Bacciagaluppi, Donald and Vermaas 
\cite{Bacciagaluppi Donald and Vermaas}, this result can be applied to the
minimal entropy proposal in cases where theorem 4 applies.

Finally, we turn to the issue of dynamics. Given theorem 4, it is possible
to show that the minimal entropy proposal satisfies the constraint of
analyticity in some cases. In particular, if the dynamical state vector
evolves in such a way that it has a bi-orthogonal decomposition with respect
to some coarse-graining of the distinguished factorization for a finite
interval of time, then Eq.(\ref{analyticity}) can be satisfied for that
interval. The reason is that the vector elements of a bi-orthogonal
decomposition are analytic functions of time, as has been shown by
Bacciagaluppi and Dickson\cite{BacciagaluppiDickson}. It remains an open
question whether for arbitrary dynamical state vectors the decomposition
that minimizes the IU entropy, considered as a function of time, satisfies
the analyticity constraint. If this is indeed the case, then the entropy
minimization rule defines a set of preferred paths.

Given such a set of paths, denoted by $\{\left| \phi _{k}(t)\right\rangle
\}_{k=1}^{d},$ we must choose the form of the dynamics from among all
possible solutions for $J_{kj}(t)$ and $T_{kj}(t)$ in Eqs. (\ref{current}), (%
\ref{antisym}) and (\ref{continuity}). We follow Bacciagaluppi and Dickson%
\cite{BacciagaluppiDickson} in choosing: 
\begin{equation}
J_{kj}(t)=2%
\mathop{\rm Im}%
\left[ \left\langle \psi (t)|\phi _{k}(t)\right\rangle \left\langle \phi
_{k}(t)|H-\tilde{H}(t)|\phi _{j}(t)\right\rangle \left\langle \phi
_{j}(t)|\psi (t)\right\rangle \right] ,  \label{chosen current}
\end{equation}
and 
\begin{equation}
T_{kj}(t)=\max \{0,\frac{J_{kj}(t)}{p_{j}(t)}\}.
\label{chosen infinitesimal parameters}
\end{equation}
This is a generalization to time-dependent preferred decompositions of the
choice made by Bell\cite{Bell}, Vink\cite{Vink} and Bub\cite{BubIQW}. Since
the inequality in Eq.(\ref{Tnm}) is saturated, this choice of $T_{kj}(t)$
minimizes the degree of stochasticity for a given form of the current. Such
a choice is motivated by the fact that classical mechanics, which is
deterministic, must be obtained as a limit of quantum mechanics.

\subsection{The quantum measurement problem}

We now consider whether the minimal entropy proposal solves the quantum
measurement problem. Although this term is often taken to refer to the whole
cluster of conceptual difficulties surrounding measurement, we shall use it
to refer to the particular problem of deriving operational quantum mechanics
from a realist no-collapse interpretation. To consider the problem, we must
introduce a quantum mechanical model of the measurement procedure, that is,
a model of the interaction between the degrees of freedom of the system
under investigation, the apparatus, and the environment. We discuss both
single measurements and sequences of measurements.

\subsubsection{Single measurements}

Following the notation introduced in section 3.1, we consider the
measurement of a Hermitian operator $V,$ belonging to a Hilbert space ${\cal %
H}^{S},$ the eigenvalues of which are non-degenerate and the eigenvectors of
which are denoted by $\{\left| \varphi _{k}\right\rangle \}_{k=1}^{m}$.
Assuming the preparation procedure is associated with a state vector $%
\sum_{k}c_{k}\left| \varphi _{k}\right\rangle ,$ where $\sum_{k}\left|
c_{k}\right| ^{2}=1$, operational quantum mechanics predicts, via the Born
rule, that the measurement will have outcome $k$ with probability $\left|
c_{k}\right| ^{2}.$

We now consider a quantum mechanical model of the measurement process.{\em \ 
}The system under investigation is called the object system and is assumed
to be microscopic. This is made to interact with a macroscopic apparatus,
associated with a Hilbert space ${\cal H}^{A},$ which in turn interacts with
a macroscopic environment, associated with a Hilbert space ${\cal H}^{E}$.
Given an initial state vector in ${\cal H}^{S}\otimes {\cal H}^{A}\otimes 
{\cal H}^{E},$ one could in principle determine the evolution of the total
system using the full microscopic Hamiltonian.

In practice of course the problem is far too complex to be solved exactly.
Nonetheless, there is a set of standard toy models of measurement that are
commonly used to investigate realist interpretations. These models adopt
some simplifying assumptions about the initial state and the form of the
evolution. Specifically, it is assumed that the object system, apparatus and
environment are all initially uncorrelated, so that the initial dynamical
state vector has the form $\left| \varphi _{k}\right\rangle \otimes \left|
A_{0}\right\rangle \otimes \left| E_{0}\right\rangle ,$ a product state with
respect to the factorization $\{{\cal H}^{S},{\cal H}^{A},{\cal H}^{E}\}$ of
the Hilbert space. The dynamics is assumed to be such that 
\begin{equation}
\left| \varphi _{k}\right\rangle \otimes \left| A_{0}\right\rangle \otimes
\left| E_{0}\right\rangle \mapsto \left| \tilde{\varphi}_{k}\right\rangle
\otimes \left| A_{k}\right\rangle \otimes \left| E_{k}\right\rangle ,
\label{eigenstate}
\end{equation}
where \{$\left| A_{k}\right\rangle \}_{k=1}^{m}$ is a set of orthonormal
vectors for the apparatus, \{$\left| E_{k}\right\rangle \}_{k=1}^{m}$ is a
set of orthonormal vectors for the environment, and \{$\left| \tilde{\varphi}%
_{k}\right\rangle \}_{k=1}^{m}$ is a set of normalized but possibly
non-orthogonal vectors for the object system, and where `$\mapsto $' denotes
the mapping corresponding to the unitary evolution.

If the initial state for the object system is $\sum_{k}c_{k}\left| \varphi
_{k}\right\rangle $, the final dynamical state vector for the total system,
given Eq.(\ref{eigenstate}) and the assumption that the evolution is linear,
is 
\begin{equation}
\left| \psi _{\text{final}}\right\rangle =\sum_{k=1}^{m}c_{k}\left| \tilde{%
\varphi}_{k}\right\rangle \otimes \left| A_{k}\right\rangle \otimes \left|
E_{k}\right\rangle .  \label{psi final}
\end{equation}
\qquad

We are now in a position to ask whether a given realist no-collapse
interpretation falls prey to the quantum measurement problem within this
model. We begin by illustrating the problem in the traditional manner,
specifically, in the context of the simplest realist no-collapse
interpretation one can imagine: one where the property ascriptions for
systems are fixed by the orthodox rule, defined in Eq.(\ref{orthodox rule}).
Such an interpretation has been called the `bare theory' by Albert\cite
{Albert}. Within the framework of section 4, it corresponds to adopting the
trivial decomposition of the dynamical state vector as preferred (the
trivial decomposition of $\left| \psi (t)\right\rangle $ is simply $%
\{(1,\left| \psi (t)\right\rangle )\}$).

Consider first a case where $c_{k}\ne 0$ for only a single value of $k,$
that is, where the initial state vector of the object system is an
eigenstate of $V.$ The final state vector is then of the form $\left| \tilde{%
\varphi}_{k}\right\rangle \otimes \left| A_{k}\right\rangle \otimes \left|
E_{k}\right\rangle $. By the orthodox rule and the reductionist rule, the
preferred projector for the property ascription to the apparatus is $%
P_{A_{k}}.$ If the bare theory is to reproduce the predictions of
operational quantum mechanics in this case, then the property associated
with the projector $P_{A_{k}}$ must be such that the apparatus can be
accurately described as `indicating outcome $k$' (for instance, if the
apparatus indicates the outcome by a digital display, $P_{A_{k}}$ could
correspond to the property of displaying the number $k$). We refer to this
as the assumption of {\em ontological correspondence.}

If, on the other hand, the initial state is such that $c_{k}\ne 0$ for more
than one value of $k,$ then the final state vector is of the form $%
\sum_{k}^{\prime }c_{k}\left| \tilde{\varphi}_{k}\right\rangle \otimes
\left| A_{k}\right\rangle \otimes \left| E_{k}\right\rangle ,$ where $%
\sum_{k}^{\prime }$ indicates a sum over values of $k$ for which $c_{k}\ne
0. $ In this case, the preferred projector for the property ascription to
the apparatus is $\sum_{k}^{^{\prime }}P_{A_{k}},$ while no projector of the
form $P_{A_{k}}$ receives the value 1. Thus, even {\em given} the assumption
of ontological correspondence, the bare theory does not predict that the
apparatus indicates the outcome $k$ for any value of $k$ for which $c_{k}\ne
0.$ Hence the bare theory does not reproduce the predictions of operational
quantum mechanics. This is the quantum measurement problem.

We now specify the assumptions under which the minimal entropy proposal
solves this problem. These involve the nature of the distinguished
factorization, which we have not yet specified. Whatever it might be, the
distinguished factorization should be defined in terms of primitives of the
theory and selected by physical principles, for instance, from
considerations of symmetry. We do not here present an argument for the
identity of the distinguished factorization, however a discussion of the
issue can be found in Dieks\cite{Dieks pref fact}, wherein it is argued that
a necessary condition on this choice is that the factor spaces carry an
irreducible representation of the space-time group (the Galilei group in
nonrelativistic quantum mechanics). For the present, we insist only that the
distinguished factorization, which we denote by $F,$ has the factorization $%
\{{\cal H}^{S},{\cal H}^{A},{\cal H}^{E}\}$ as a coarse-graining, and that
its elements correspond to microscopic degrees of freedom (for instance,
they could correspond to degrees of freedom of elementary particles).

Now, suppose that $F$ is such that all the vectors in the sets $\{\left| 
\tilde{\varphi}_{k}\right\rangle \}_{k=1}^{m},$ $\{\left| A_{k}\right\rangle
\}_{k=1}^{m},$ and $\{\left| E_{k}\right\rangle \}_{k=1}^{m}$ are product
states with respect to it$.$ One can then determine that the preferred
decomposition of $\left| \psi _{\text{final}}\right\rangle $ is 
\[
D_{\text{final}}=\left\{ \left( c_{k},\left| \tilde{\varphi}%
_{k}\right\rangle \otimes \left| A_{k}\right\rangle \otimes \left|
E_{k}\right\rangle \right) \right\} _{k=1}^{m}. 
\]
This follows from theorem 4 and the fact that $D_{\text{final}}$ is a
product decomposition with respect to $F$ that is bi-orthogonal with respect
to the coarse-graining $F_{\text{bi}}=\{{\cal H}^{S}\otimes {\cal H}^{A},%
{\cal H}^{E}\}$ of $F.$ It then follows from the Born rule constraint that,
with probability $\left| c_{k}\right| ^{2},$ the property state vector is 
\[
\left| \Phi _{\text{final}}\right\rangle =\left| \tilde{\varphi}%
_{k}\right\rangle \otimes \left| A_{k}\right\rangle \otimes \left|
E_{k}\right\rangle . 
\]
Using the reductionist rule we find that the projector $P_{A_{k}}$ is
determinate and receives the value $1$ with probability $\left| c_{k}\right|
^{2}.$ Finally, by the assumption of ontological correspondence, the
apparatus has the property of indicating outcome $k$ with probability $%
\left| c_{k}\right| ^{2}.$ This is in agreement with the predictions of
operational quantum mechanics.

Thus, we have obtained a solution to the measurement problem within the
standard model of measurement. In so doing, we have had to assume that when
the initial state vector of the object system is an eigenstate of $V,$ the
final dynamical state vector for the total system is unentangled with
respect to the distinguished factorization.

The assumption of no entanglement between the distinguished factor spaces of
the apparatus and the environment is not particularly realistic, given that
these factor spaces are taken to correspond to microscopic degrees of
freedom, and typical interactions between the apparatus and the environment
are likely to entangle these degrees of freedom. However, this assumption
can be relaxed somewhat without changing any of our conclusions, as we now
demonstrate.

We consider a model of measurement wherein the evolution is of the form: 
\begin{equation}
\left| \varphi _{k}\right\rangle \otimes \left| A_{0}\right\rangle \otimes
\left| E_{0}\right\rangle \mapsto \left| \tilde{\varphi}_{k}\right\rangle
\otimes \sum_{\mu =1}^{M}f_{k,\mu }\left( \left| A_{k,\mu }\right\rangle
\otimes \left| E_{k,\mu }\right\rangle \right) .  \label{eigenstate 2}
\end{equation}
where $\sum_{\mu }\left| f_{k,\mu }\right| ^{2}=1$ and where $\{\{\left|
A_{k,\mu }\right\rangle \}_{\mu =1}^{M}\}_{k=1}^{m}$ and $\{\{\left|
E_{k,\mu }\right\rangle \}_{\mu =1}^{M}\}_{k=1}^{m}$ are orthonormal sets of
vectors that are product states with respect to $F.$ The assumption of
ontological correspondence in this case becomes the assumption that for
every value of $\mu ,$ the projector $P_{A_{k,\mu }}$ corresponds to the
apparatus indicating outcome $k$. That there can be more than one projector
corresponding to indicating a particular outcome is not unreasonable since
there can be many different microscopic configurations of the apparatus
leading to the same overall macroscopic appearance.

An arbitrary initial state vector for the object system, $%
\sum_{k}c_{k}\left| \varphi _{k}\right\rangle ,$ leads, via Eq.(\ref
{eigenstate 2}), to the following final dynamical state vector for the total
system 
\[
\left| \psi _{\text{final}}^{\prime }\right\rangle
=\sum_{k=1}^{m}c_{k}\left| \tilde{\varphi}_{k}\right\rangle \otimes
\sum_{\mu =1}^{M}f_{k,\mu }\left( \left| A_{k,\mu }\right\rangle \otimes
\left| E_{k,\mu }\right\rangle \right) . 
\]
\qquad The preferred decomposition of this state vector is 
\[
D_{\text{final}}^{\prime }=\left\{ \left\{ \left( c_{k}f_{k,\mu },\left| 
\tilde{\varphi}_{k}\right\rangle \otimes \left| A_{k,\mu }\right\rangle
\otimes \left| E_{k,\mu }\right\rangle \right) \right\} _{\mu
=1}^{M}\right\} _{k=1}^{m}, 
\]
since this is a product decomposition with respect to $F$ that is
bi-orthogonal with respect to $F_{\text{bi}}.$ It follows that the property
state vector is 
\[
\left| \Phi _{\text{final}}^{\prime }\right\rangle =\left| \tilde{\varphi}%
_{k}\right\rangle \otimes \left| A_{k,\mu }\right\rangle \otimes \left|
E_{k,\mu }\right\rangle 
\]
with probability $\left| c_{k}f_{k,\mu }\right| ^{2}.$ By the reductionist
rule, the projector $P_{A_{k,\mu }}$ is determinate and receives value $1$
with probability $\left| c_{k}f_{k,\mu }\right| ^{2}.$ Finally, by the
assumption of ontological correspondence, the apparatus has the property of
indicating outcome $k$ with probability $\sum_{\mu }\left| c_{k}f_{k,\mu
}\right| ^{2}=\left| c_{k}\right| ^{2},$ in agreement with operational
quantum mechanics.

Note that the model of measurement provided by Eq.(\ref{eigenstate 2}) can
also describe error-prone measurements. This occurs if for some values of $%
\mu ,$ $P_{A_{k,\mu }}$ corresponds to the property of indicating an outcome 
$k^{\prime }\ne k,$ or to the property of indicating a malfunction.
Furthermore, this model can incorporate measurements described by positive
operator-valued measures(POVMs)\cite{Peres}. This follows from the fact that
such measurements are implemented by adjoining an ancilla to the system
under investigation and measuring a projector-valued measure(PVM) on the
composite. By including the ancilla in our definition of the object system,
the model presented above can describe these measurements. Note however that
we are restricted to PVMs whose eigenvectors are product states with respect
to the distinguished factorization.

Despite the possibility of incorporating some error-prone and POVM
measurements, the model of measurement provided by Eq.(\ref{eigenstate 2})
is still not the most general or realistic. Although it is true that an
arbitrary state vector has many decompositions into product states with
respect to the distinguished factorization, it is not necessarily the case
that any of these decompositions are bi-orthogonal with respect to a
coarse-graining of the distinguished factorization. For instance, if any of
the vectors in the set $\{\left| \tilde{\varphi}_{k}\right\rangle
\}_{k=1}^{m}$ are entangled with respect to the distinguished factor spaces
of ${\cal H}^{S},$ then theorem 4 fails to apply if the final dynamical
state vector is of the form of $\left| \psi _{\text{final}}^{\prime
}\right\rangle $. Since the problem of minimizing the IU entropy for
arbitrary state vectors has not yet been solved, it is not clear what the
preferred decomposition will be in this case and whether the measurement
problem is resolved or not.

It is nonetheless interesting to consider one particular type of
modification of the evolution where the only change from the model
considered above is that the set of vectors $\{\{\left| E_{k,\mu
}\right\rangle \}_{\mu =1}^{M}\}_{k=1}^{m}$ (describing the states of the
environment that are relative to the apparatus states $\{\{\left| A_{k,\mu
}\right\rangle \}_{\mu =1}^{M}\}_{k=1}^{m}$) is only approximately
orthogonal. This is an instance where theorem 4 may fail to apply. However,
the difference between $\left| \psi _{\text{final}}^{\prime }\right\rangle $
when the elements of $\{\{\left| E_{k,\mu }\right\rangle \}_{\mu
=1}^{M}\}_{k=1}^{m}$ are orthogonal and when they are very nearly
orthogonal, is not significant. Thus, if the preferred decomposition does
not depend sensitively on small variations in the dynamical state vector,
the preferred decomposition in the nearly orthogonal case should be `close
to' $D_{\text{final}}^{\prime },$ and it is then likely that the apparatus
will be assigned an ontology that is `close to' the one it receives for the
orthogonal case. We see therefore that whether or not there is a measurement
problem in this case depends on whether or not there is such sensitive
dependence. The answer to this question must await further progress on the
problem of the minimization of the IU entropy.\footnote{%
The analagous question in the Vermaas-Dieks version of the modal
interpretation is whether the spectral resolution of a density operator is
sensitive to small changes in the density operator. Bacciagaluppi, Donald
and Vermaas\cite{Bacciagaluppi Donald and Vermaas} have shown that this does
in fact occur when the density operator has nearly degenerate eigenvalues.}

\strut Finally, we note that the assumption that the apparatus and
environment are {\em initially} unentangled is also an unrealistic feature
of the standard model of measurement. For that matter, the assumption that
the composite of system, apparatus and environment is unentangled with the
rest of the universe may not be realistic either.{\em \ }However, this
difficulty is not unique to the minimal entropy proposal. Every realist
no-collapse interpretation must contend with the fact that the dynamical
state vector for the universe is in general not factorizable with respect to
subsystems that have interacted in the past, even if this interaction is
quite weak. Further work is required to determine whether the predictions of
the minimal entropy proposal remain satisfactory when these assumptions are
relaxed.

\subsubsection{Sequences of measurements}

We now demonstrate the extent to which the minimal entropy proposal is in
agreement with operational quantum mechanics for {\em sequences} of
measurements. Consider in particular the sequence of two measurements
described in section 3.1. Recall that the first measurement is of a variable 
$V$ with eigenstates $\{\left| \varphi _{k}\right\rangle \}_{k=1}^{m}$, the
second measurement is of a variable $V^{\prime }$ with eigenstates $\{\left|
\varphi _{k}^{\prime }\right\rangle \}_{k=1}^{m},$ and the state prepared by
the first apparatus given outcome $k$ is denoted by $\left| \tilde{\varphi}%
_{k}\right\rangle .$

In the last subsection we considered two distinct models of measurement
which differed in the extent to which the apparatus and the environment
became entangled due to their interaction. In this subsection, we consider
only the simpler of the two models. The reader can verify that the more
realistic model leads to the same conclusions.

There are now two apparatuses, and an environment for each. We denote their
Hilbert spaces by ${\cal H}^{A1},{\cal H}^{A2},{\cal H}^{E1},$ and ${\cal H}%
^{E2}$ respectively, and we distinguish state vectors for the two
apparatuses(environments) by a superscript. It is again assumed that the
object system, the two apparatuses and the two environments are all
initially uncorrelated. The distinguished factorization $F$ is assumed to
have $\{{\cal H}^{S},{\cal H}^{A1},{\cal H}^{A2},{\cal H}^{E1},$ ${\cal H}%
^{E2}\}$ as a coarse-graining.

We assume that the first measurement is well described by Eq.(\ref
{eigenstate}) with the exception of a change of notation: $\left|
A_{0}\right\rangle ,\left| E_{0}\right\rangle ,\left| A_{k}\right\rangle $
and $\left| E_{k}\right\rangle $ become $\left| A_{0}^{1}\right\rangle
,\left| E_{0}^{1}\right\rangle ,\left| A_{k}^{1}\right\rangle $ and $\left|
E_{k}^{1}\right\rangle $ in order to specify that the object system
interacts with the first rather than the second apparatus. We assume that
the second apparatus and its environment remain uncorrelated with the rest
of the system and each other during this first measurement$.${\em \ }It
follows that the dynamical state vector for the total system after the first
measurement is 
\begin{equation}
\left| \psi _{\text{final 1}}\right\rangle =\left( \sum_{k=1}^{m}c_{k}\left| 
\tilde{\varphi}_{k}\right\rangle \otimes \left| A_{k}^{1}\right\rangle
\otimes \left| E_{k}^{1}\right\rangle \right) \otimes \left|
A_{0}^{2}\right\rangle \otimes \left| E_{0}^{2}\right\rangle .
\label{psi final 1}
\end{equation}

Suppose that $\left| A_{0}^{2}\right\rangle $ and $\left|
E_{0}^{2}\right\rangle $ are product states with respect to $F.$ If we make
all the same assumptions about $\left| A_{k}^{1}\right\rangle $ and $\left|
E_{k}^{1}\right\rangle $ as were made for $\left| A_{k}\right\rangle $ and $%
\left| E_{k}\right\rangle ${\em \ }in the previous subsection, and if we use
the bi-partite factorization $\{{\cal H}^{S}\otimes {\cal H}^{A1}\otimes 
{\cal H}^{A2},{\cal H}^{E1}\otimes {\cal H}^{E2}\}$ in place of the
bi-partite factorization $\{{\cal H}^{S}\otimes {\cal H}^{A},{\cal H}^{E}\}$
in the arguments found therein, then it is straightforward to show that the
preferred decomposition of $\left| \psi _{\text{final 1}}\right\rangle $ is 
\begin{equation}
D_{\text{final 1}}=\left\{ \left( c_{k},\left| \tilde{\varphi}%
_{k}\right\rangle \otimes \left| A_{k}^{1}\right\rangle \otimes \left|
E_{k}^{1}\right\rangle \otimes \left| A_{0}^{2}\right\rangle \otimes \left|
E_{0}^{2}\right\rangle \right) \right\} _{k=1}^{m}.  \label{D final 1}
\end{equation}
We conclude that with probability $\left| c_{k}\right| ^{2},$ the first
apparatus indicates outcome $k,$ while the second apparatus remains ready to
measure.

Now assume that the second measurement is also well described by Eq.(\ref
{eigenstate}) with the notational change that $\left| A_{0}\right\rangle
,\left| E_{0}\right\rangle ,\left| A_{k}\right\rangle $ and $\left|
E_{k}\right\rangle $ become $\left| A_{0}^{2}\right\rangle ,\left|
E_{0}^{2}\right\rangle ,\left| A_{k}^{2}\right\rangle $ and $\left|
E_{k}^{2}\right\rangle $ since the object system is now interacting with the
second apparatus, and where the vectors for the object system acquire a
prime since the second measurement is of $V^{\prime }$ rather than $V.$ For
simplicity, we take this second measurement to be non-disturbing, so that $%
\left| \tilde{\varphi}_{k}^{\prime }\right\rangle =\left| \varphi
_{k}^{\prime }\right\rangle .$ Assume also that the first apparatus and its
environment have no interactions during this measurement. It then follows
that the dynamical state vector after the second measurement is 
\begin{equation}
\left| \psi _{\text{final 2}}\right\rangle
=\sum_{k=1}^{m}\sum_{j=1}^{m}c_{k}d_{j}^{k}\left| \varphi _{j}^{\prime
}\right\rangle \otimes \left| A_{k}^{1}\right\rangle \otimes \left|
E_{k}^{1}\right\rangle \otimes \left| A_{j}^{2}\right\rangle \otimes \left|
E_{j}^{2}\right\rangle ,  \label{psi final 2}
\end{equation}
where the coefficients $\{d_{j}^{k}\}_{k=1}^{m}$ are defined by $\left| 
\tilde{\varphi}_{k}\right\rangle =\sum_{k=1}^{m}d_{j}^{k}\left| \varphi
_{j}^{\prime }\right\rangle .$ Again, if we make all the same assumptions
about $\left| A_{j}^{2}\right\rangle $ and $\left| E_{j}^{2}\right\rangle $
as were made for $\left| A_{j}\right\rangle $ and $\left| E_{j}\right\rangle 
$ in the previous subsection, and if we use the bi-partite factorization $\{%
{\cal H}^{S}\otimes {\cal H}^{A1}\otimes {\cal H}^{A2},{\cal H}^{E1}\otimes 
{\cal H}^{E2}\}$ in place of the bi-partite factorization $\{{\cal H}%
^{S}\otimes {\cal H}^{A},{\cal H}^{E}\}$ in all the arguments found therein,
the preferred decomposition of $\left| \psi _{\text{final 2}}\right\rangle $
is found to be 
\begin{equation}
D_{\text{final 2}}=\left\{ \left\{ \left( c_{k}d_{j}^{k},\left| \varphi
_{j}^{\prime }\right\rangle \otimes \left| A_{k}^{1}\right\rangle \otimes
\left| E_{k}^{1}\right\rangle \otimes \left| A_{j}^{2}\right\rangle \otimes
\left| E_{j}^{2}\right\rangle \right) \right\} _{k=1}^{m}\right\} _{j=1}^{m}.
\label{D final 2}
\end{equation}
We can therefore conclude that there is a probability $\left|
c_{k}d_{j}^{k}\right| ^{2}$ that the first apparatus indicates outcome $k$
and the second apparatus indicates outcome $j$ after the second measurement.
It follows that the probability of the second apparatus indicating outcome $%
j $ {\em given} that the first apparatus indicates outcome $k$ after the
second measurement is $\left| d_{j}^{k}\right| ^{2}=\left| \left\langle
\varphi _{j}|\tilde{\varphi}_{k}\right\rangle \right| ^{2}.$

However, we have still not determined the probability for the second
apparatus to indicate outcome $j$ after the second measurement given that
the first apparatus indicates outcome $k$ {\em after the first measurement},
which is the quantity specified by the generalized Born rule. The problem is
that it has not been shown that the outcome indicated by the first apparatus
is stable over time. Whether it is or not depends on the dynamics of the
property state vector, which is determined by Eqs. (\ref{chosen current})
and (\ref{chosen infinitesimal parameters}). Now although it may be
reasonable to assume that the dynamical state vector after a measurement is
such that theorem 4 applies, it is unlikely that this theorem applies during
the entire interaction leading up to this outcome. Given this, we cannot at
present determine the time-sequence of preferred decompositions nor the
preferred paths through Hilbert space defined by this sequence. Since Eqs. (%
\ref{chosen current}) and (\ref{chosen infinitesimal parameters}) depend on
the identity of these preferred paths, we cannot at present determine the
dynamics of the property state vector.

Thus, for the moment we simply {\em assume} that within the minimal entropy
proposal, the apparatus is never described as `jumping' between
macroscopically different readings. We call this assumption {\em stability. }%
Given stability, the minimal entropy proposal reproduces the predictions of
the generalized Born rule.

\subsection{The Faithfulness criterion revisited}

We now reconsider the experiment of section 3.1 in the context of the
minimal entropy proposal. Since this experiment involves a sequence of two
measurements, we can make use of the model of measurement presented in the
previous section. Consider the property ascription at the time $t,$ after
the first measurement. The dynamical state vector is $\left| \psi _{\text{%
final 1}}\right\rangle $, defined in Eq.(\ref{psi final 1}), and its
preferred decomposition is $D_{\text{final 1}},$ defined in Eq.(\ref{D final
1}). If the first apparatus indicates the outcome $k$ at time $t$, the
property state vector must be the $k$th element of $D_{\text{final 1}},$
that is, $\left| \Phi _{\text{final 1}}\right\rangle =\left| \tilde{\varphi}%
_{k}\right\rangle \otimes \left| A_{k}^{1}\right\rangle \otimes \left|
E_{k}^{1}\right\rangle \otimes \left| A_{0}^{2}\right\rangle \otimes \left|
E_{0}^{2}\right\rangle .$ The critical feature of the experiment of section
3.1 is that the vector $\left| \tilde{\varphi}_{k}\right\rangle $ that is
prepared when the first apparatus indicates outcome $k,$ is an eigenstate of 
$V_{(k)}$, the variable measured by the second apparatus. It follows that
the variable $V_{(k)}\otimes I$ (where $I$ is the identity operator for $%
{\cal H}^{A1}\otimes {\cal H}^{A2}\otimes {\cal H}^{E1}\otimes {\cal H}%
^{E2}) $ is determinate and has value $v_{(k),1}$ at time $t.$ Finally, it
follows from the reductionist rule that $V_{(k)}$ is determinate and has
value $v_{(k),1}$ at time $t.$ This is precisely what is required in order
for the faithfulness criterion to be satisfied.

It should be noted that since the variable measured by the second apparatus
depends upon the outcome of the first measurement, the initial state of the
second apparatus may well be different for different outcomes of the first
measurement. Thus, rather than the first measurement interaction being
described by Eq.(\ref{eigenstate}), it may be described by 
\begin{eqnarray}
&&\left| \varphi _{k}\right\rangle \otimes \left| A_{0}^{1}\right\rangle
\otimes \left| E_{0}^{1}\right\rangle \otimes \left| A_{0}^{2}\right\rangle
\otimes \left| E_{0}^{2}\right\rangle  \nonumber \\
&\mapsto &\left| \tilde{\varphi}_{k}\right\rangle \otimes \left|
A_{k}^{1}\right\rangle \otimes \left| E_{k}^{1}\right\rangle \otimes \left|
A_{(k),0}^{2}\right\rangle \otimes \left| E_{(k),0}^{2}\right\rangle ,
\end{eqnarray}
where $\{\left| A_{(k),0}^{2}\right\rangle \}_{k=1}^{m}$ and $\{\left|
E_{(k),0}^{2}\right\rangle \}_{k=1}^{m}$ are orthonormal sets of vectors,
and $\left| A_{(k),0}^{2}\right\rangle $ corresponds to the apparatus being
ready to measure the variable $V_{(k)}.$ In any event, by making the same
assumptions for $\left| A_{(k),0}^{2}\right\rangle $ and $\left|
E_{(k),0}^{2}\right\rangle $ as were made for $\left| A_{0}^{2}\right\rangle 
$ and $\left| E_{0}^{2}\right\rangle $ in the last subsection, one can show
that the minimal entropy proposal is in agreement with the predictions of
operational quantum mechanics even when the nature of the second measurement
depends on the outcome of the first.

We end this section with a discussion of the case wherein the second
measurement is of a variable whose eigenstates are not all product states
with respect to the distinguished factorization $F$. For such measurements,
the faithfulness criterion cannot be satisfied within the minimal entropy
proposal. The reason is as follows. Suppose $\left| \varphi \right\rangle $
is an eigenstate of the measured variable that is entangled with respect to $%
F.$ If $\left| \varphi \right\rangle $ is prepared by the first measurement
and measured by the second, then the faithfulness criterion requires that
the projector $P_{\varphi }$ be determinate with value $1$ immediately prior
to the second measurement. However, for this to occur the property state
vector must be an eigenstate of $P_{\varphi },$ and hence must be entangled
with respect to $F$. But the property state vector is always a product state
with respect to $F$ in the minimal entropy proposal.

The failure of the faithfulness criterion for such measurements in the
context of the atomic modal interpretation of Bacciagaluppi and Dickson\cite
{BacciagaluppiDickson} and Dieks\cite{Dieks pref fact} has been discussed by
Dieks, and also by Vermaas\cite{Vermaas in pink book}. These authors have
suggested that an explanation of the outcomes of these measurements might be
provided by dispositional properties or collective effects of the composite.
This explanation can also be invoked in the context of the minimal entropy
proposal.

\section{\strut Conclusions}

In modal interpretations, the properties of a system are given by a
specification of the set of determinate variables (the ontology) and the
value ascription to these variables, jointly referred to as the property
ascription. Such interpretations also assume that the property ascription
which obtains at a given time is just one of several possibilities. There is
always a unique `smallest' projector which receives the value $1$ in each of
these possible property ascriptions, which we call the preferred projector
for that property ascription.

We have shown that these preferred projectors must be non-orthogonal if one
seeks to satisfy the faithfulness criterion, that is, if one seeks to
explain the outcomes of certain perfectly predictable measurements in terms
of pre-existing properties of the system under investigation. The
possibility of such an explanation has historically been a strong motivation
for the modal approach.

We have also shown that non-orthogonal preferred projectors are inconsistent
with the assumption, common among previous modal interpretations, that at a
given time there is only a single possible ontology. In order to consider
non-orthogonal preferred projectors, we have developped a framework for
modal interpretations wherein at a given time, the possible property
ascriptions may differ with respect to ontology. As is required for any
modal interpretation, the state vector appearing in the Schr\"{o}dinger
equation, which we call the dynamical state vector, does not uniquely fix
the property ascription. Rather, a preferred decomposition of the dynamical
state vector into a sum of orthogonal vectors must be specified at every
time, and a single element of this decomposition, dubbed the property state
vector, fixes the property ascription. The property state vector evolves
stochastically according to a Markovian dynamics. Finally, subsystems
receive only those properties they inherit from the total system by the
reductionist rule.

It is of course possible to generalize this framework in many ways. One
could consider non-Markovian dynamics, alternatives to the reductionist
rule, and even non-orthogonal decompositions of the dynamical state vector.
Nonetheless, we feel that the framework presented is a natural starting
place for the interpretive program at hand.

Within the context of this framework, we have presented a novel proposal for
the preferred decomposition. The proposal assumes that there is a
distinguished set of subsystems of the universe, that is, a distinguished
factorization of the total Hilbert space into a tensor product of Hilbert
spaces\footnote{%
Other modal interpreters\cite{Cliftonproperties}\cite{Vermaasno-go}\cite
{BacciagaluppiDickson} have been led to this assumption by considerations of
the correlations between the properties of a system and its subsystems.}. It
is also assumed that the preferred decomposition is a product decomposition
with respect to this factorization. In the case of a distinguished
factorization that is bi-partite, it is then natural to follow previous
authors in identifying the bi-orthogonal decomposition as preferred.{\em \ }%
However, the obvious generalization of the bi-orthogonal decomposition to an 
$n$-partite distinguished factorization, namely the $n$-orthogonal
decomposition, does not exist for all state vectors, as shown by Peres\cite
{Peres n-orthogonal}. The preferred decomposition in our proposal is the one
that minimizes the IU entropy from among all product decompositions with
respect to the distinguished factorization. This decomposition always exists
and turns out to be equal to the $n$-orthogonal decomposition when the
latter exists. It therefore can be thought of as a natural generalization of
the bi-orthogonal decomposition to $n$-partite systems.

At present the strongest justification for the minimal entropy proposal is
its success in dealing with the quantum measurement problem and in
satisfying the faithfulness criterion. The measurement problem is resolved
for a wide variety of measurements including certain types of non-ideal
measurements, in particular, disturbing measurements, assuming particular
microscopic models of the apparatus and environment. Within the same
microscopic models, the faithfulness criterion is satisfied for sequences of
disturbing measurements. It is this feature of the minimal entropy proposal
that sets it apart from previous modal interpretations.

The solution of the measurement problem relies on the assumption of
ontological correspondence, that the ontology of macroscopic systems
corresponds to our everyday perceptions of them, and the assumption of
stability, that the dynamics of the properties assigned to macroscopic
systems are consistent with our stable perceptions of them. Ideally, these
features would be demonstrated rather than assumed. However, the
demonstration of ontological correspondence is likely to require a better
specification of the distinguished factorization than has been provided in
the present work, while the demonstration of stability must await progress
in solving the entropy minimization problem in cases where theorem 4 does
not apply. Progress on the minimization problem will also help to determine
whether one can solve the quantum measurement problem for more general types
of measurements than the ones considered here, for example, measurements of
variables whose eigenstates are entangled with respect to the distinguished
factorization. In addition, such progress is required to determine what the
proposal has to say about more realistic models of measurements. Finally, it
may indicate whether the IU entropy is the correct quantity to minimize in
the rule for determining the preferred decomposition, or whether some other
entropic quantity might be a better choice.

So we see that there remain many unanswered questions. In addition to these,
there are difficulties with the minimal entropy proposal. For one, the
product decomposition that minimizes the IU entropy may fail to be unique
for certain dynamical state vectors. It may be that further technical work
will show that this is not a problem after all. For instance, dynamical
state vectors for which the preferred decomposition is non-unique for a
finite interval of time may constitute a set of measure zero. Another
difficulty is that the faithfulness criterion explicitly fails to be
satisfied in measurements of variables whose eigenstates are entangled with
respect to the distinguished factorization. Given that not all Hermitian
operators necessarily correspond to variables that can be measured\cite
{Araki and Yanase}\cite{Omnes}, it may happen that with a suitable choice of
distinguished factorization, the measurements for which the faithfulness
criterion fails to be satisfied are precisely those which are impossible to
implement. On the other hand, it may be that this problem cannot be avoided
within the minimal entropy proposal, but can be avoided if some other choice
of preferred decomposition is made. As a third possibility, one might find
that the faithfulness criterion for variables with entangled eigenstates,
cannot be satisfied by {\em any} interpretation within the framework we have
set out. Justifying any one of these answers would certainly be an
interesting result, and motivates further investigation of these issues.

The use of a preferred decomposition, sometimes called an `interpretation
basis', has been viewed by some as necessary within interpretive strategies
distinct from modal interpretations. This has been suggested by Deutsch\cite
{Deutsch} in the context of the many-worlds interpretation and by Kent and
McElwaine\cite{Kent and McElwaine} in the context of consistent histories. A
preferred decomposition might also be useful in nonlinear modifications of
quantum mechanics. Thus, the preferred decomposition of the minimal entropy
proposal may well be of relevance to such interpretive strategies as well.
In any event, a mathematically precise proposal, even though not without
problems, can be useful in stimulating progress on interpretive issues, as
is evidenced by the recent profusion of work on modal interpretations. We
hope that the minimal entropy proposal will not be an exception in this
respect.

\section{Acknowledgments}

We wish to thank Rob Clifton for helpful comments on a draft of this paper.
This work was supported by the National Sciences and Engineering Research
Council of Canada.

\section{Appendix: Proof of theorem 3.}

It will be assumed throughout that the distinguished factorization is
bi-partite, and the two factor spaces are denoted by ${\cal H}^{A}$ and $%
{\cal H}^{B}.$ All references to product decompositions are to be understood
as product decompositions with respect to this factorization. We say that a
product decomposition $\{(c_{k},\left| \chi _{k}^{A}\right\rangle \otimes
\left| \phi _{k}^{B}\right\rangle )\}_{k=1}^{m}$ is $A${\it -orthogonal} $(B$%
{\it -orthogonal}) if the set of vectors $\{\left| \chi
_{k}^{A}\right\rangle \}_{k=1}^{m}$ $\left( \{\left| \phi
_{k}^{B}\right\rangle \}_{k=1}^{m}\right) $is orthogonal. A {\it %
bi-orthogonal }decomposition (also called a Schmidt decomposition) is one
that is both $A$-orthogonal and $B$-orthogonal. We shall make use of several
well-known properties of bi-orthogonal decompositions, an exposition of
which can be found in Ref. \cite{Hughston Josza and Wooters}. Finally, we
remind the reader that $S_{\left| \psi \right\rangle }(D)$ denotes the IU
entropy of $\left| \psi \right\rangle $ for the decomposition $D,$ which is
defined by Eq.(\ref{ontological entropy}).

Theorem 3 follows from two lemmas:

\begin{description}
\item[Lemma A.1]  \ For any vector $\left| \psi \right\rangle ,$ if $D$ is
an arbitrary product decomposition of $\left| \psi \right\rangle ,$ then
there always exists an $A$-orthogonal decomposition of $\left| \psi
\right\rangle ,$ $D_{\text{A-orth}}$, such that 
\[
S_{\left| \psi \right\rangle }(D_{\text{A-orth}})\le S_{\left| \psi
\right\rangle }(D).
\]

\item[Lemma A.2]  \ For any vector $\left| \psi \right\rangle ,$ if $D_{A%
\text{-orth}}$ is any $A$-orthogonal decomposition of $\left| \psi
\right\rangle ,$ and $D_{\text{bi-orth}}$ is any bi-orthogonal decomposition
of $\left| \psi \right\rangle ,$ then 
\[
S_{\left| \psi \right\rangle }(D_{\text{bi-orth}})\le S_{\left| \psi
\right\rangle }(D_{A-\text{orth}}). 
\]
\end{description}

Together these imply that for any vector $\left| \psi \right\rangle ,$ if $D$
is an arbitrary product decomposition of $\left| \psi \right\rangle $, and $%
D_{\text{bi-orth}}$ is any bi-orthogonal decomposition of $\left| \psi
\right\rangle ,$ then 
\[
S_{\left| \psi \right\rangle }(D_{\text{bi-orth}})\le S_{\left| \psi
\right\rangle }(D), 
\]
which is simply theorem 3.

\strut The task at hand is therefore to prove lemmas A.1 and A.2. \strut We
begin by reviewing a partial ordering relation among probability
distributions, namely that of {\it majorization}\cite{Bhatia}{\it , }which
has recently seen application in the study of entanglement purification\cite
{Nielson}. Suppose ${\bf p}\equiv (p_{1},p_{2},...,p_{m})$ and ${\bf q}%
\equiv (q_{1},q_{2},...,q_{m})$ are two $m$-element probability
distributions. By definition, ${\bf p}$ {\it majorizes} ${\bf q}$ if for
every $l$ in the range $\{1,..,m\},$ 
\[
\sum_{k=1}^{l}p_{k}^{\downarrow }\ge \sum_{k=1}^{l}q_{k}^{\downarrow }, 
\]
where $p_{k}^{\downarrow }$ indicates the $k$th largest element of ${\bf p,}$
so that $p_{1}^{\downarrow }\ge p_{2}^{\downarrow }\ge ...\ge
p_{m}^{\downarrow }.$

The notion of majorization is important in the present investigation because
of the following well-known result (Theorem II.3.1 of Ref. \cite{Bhatia}):
The following two conditions are equivalent 
\begin{eqnarray*}
&&(i)\text{ }{\bf p}\text{ majorizes }{\bf q.} \\
(ii)\text{ }\sum_{k=1}^{m}f(p_{k}) &\le &\sum_{k=1}^{m}f(q_{k})\text{ for
all concave functions }f.
\end{eqnarray*}
Since $-x\log x$ is a concave function of $x$, it follows that $H({\bf p)\le 
}H({\bf q)}$ if and only if ${\bf p}$ majorizes ${\bf q,}$ where $H({\bf p)}$
is the Shannon entropy of a probability distribution ${\bf p}$, defined in
Eq.(\ref{shannonentropy}). Now consider two decompositions of a state
vector, $D=\{(c_{k},\left| \phi _{k}\right\rangle \}_{k=1}^{m}$ and $%
D^{\prime }=\{(c_{k}^{\prime },\left| \phi _{k}^{\prime }\right\rangle
\}_{k=1}^{m^{\prime }}.$ Although these may have different cardinalities,
they can be associated with probability distributions of equal cardinality
by simply adding zeroes. Specifically, if $m\ge m^{\prime },$ then $D$ is
associated with the distribution $p_{k}=\left| c_{k}\right| ^{2}$ for $k\in
\{1,...,m\}$ and $D^{\prime }$ is associated with the distribution $q_{k}=$ $%
\left| c_{k}^{\prime }\right| ^{2}$ for $k\in \{1,...,m^{\prime }\}$ and $%
q_{k}=0$ for $k\in \{m^{\prime }+1,...,m\}.$ Since the IU entropy of $\left|
\psi \right\rangle $ for the decomposition $D(D^{\prime })$ is simply the
Shannon entropy of ${\bf p}$(${\bf q)}$, it follows that $S_{\left| \psi
\right\rangle }(D)\le S_{\left| \psi \right\rangle }(D^{\prime })$ if ${\bf p%
}$ majorizes ${\bf q.}$

In order to facilitate the proof of lemma A.1, we set out two minor lemmas.

\begin{description}
\item[Lemma A.3]  \ Consider two probability distributions ${\bf p\equiv (}%
p_{1},p_{2},...,p_{m})$ and ${\bf q\equiv (}q_{1},q_{2},...,q_{m}).$ If for
every $l$ in the range $\{1,...,m\},$%
\[
\sum_{k=1}^{l}p_{_{k}}\ge \sum_{k=1}^{l}q_{k}^{\downarrow }, 
\]
then ${\bf p}$ majorizes ${\bf q.}$
\end{description}

{\bf Proof.} This result follows from the definition of majorization and the
fact that 
\[
\sum_{k=1}^{l}p_{k}^{\downarrow }\ge \sum_{k=1}^{l}p_{k} 
\]
for every $l$ in the range $\{1,...,m\}{\bf .}$ This inequality is obviously
true since the $l$-element subset of ${\bf p}$ with the largest sum must be
the subset containing the $l$ largest elements of ${\bf p.}$ {\bf QED.}

For the second minor lemma, we make use of some notational conventions
introduced in the text: $P_{{\cal S}}$ denotes the projector onto the
subspace ${\cal S},$ and `$P_{{\cal S}}<P_{{\cal S}^{\prime }}$' denotes
that ${\cal S}$ is a proper subspace of ${\cal S}^{\prime }.$

\begin{description}
\item[Lemma A.4]  \ If $P_{{\cal S}}\le P_{{\cal S}^{\prime }}$ then $%
\left\langle \psi \right| P_{{\cal S}}\left| \psi \right\rangle \le
\left\langle \psi \right| P_{{\cal S}^{\prime }}\left| \psi \right\rangle .$
\end{description}

{\bf Proof.} If $P_{{\cal S}}=P_{{\cal S}^{\prime }},$ then the inequality
is saturated. Otherwise, $P_{{\cal S}}<P_{{\cal S}^{\prime }},$ and there
exists a projector $P_{{\cal T}}$ such that $P_{{\cal S}}+P_{{\cal T}}=P_{%
{\cal S}^{\prime }}.$ The desired inequality follows from the positivity of $%
\left\langle \psi \right| P_{{\cal T}}\left| \psi \right\rangle .$ {\bf QED.}

We are now in a position to prove lemma A.1.

{\bf Proof of lemma A.1.} An arbitrary product decomposition has the form $%
D=\{(d_{k},\left| \phi _{k}^{A}\right\rangle \otimes \left| \chi
_{k}^{B}\right\rangle )\}_{k=1}^{m},$ where the lists of vectors $\{\left|
\phi _{k}^{A}\right\rangle \}_{k=1}^{m}$ and $\{\left| \chi
_{k}^{A}\right\rangle \}_{k=1}^{m}$ are not necessarily orthogonal nor even
linearly independent (although the list of vectors $\{\left| \phi
_{k}^{A}\right\rangle \otimes \left| \chi _{k}^{B}\right\rangle \}_{k=1}^{m}$
{\em is} orthogonal)$.$ The decomposition $D$ defines an $m$-element
probability distribution ${\bf q=\{}q_{1},q_{2},...,q_{m}\},{\bf \ }$where $%
q_{k}\equiv \left| d_{k}\right| ^{2}.$ As before, let $q_{k}^{\downarrow }$
denote the $k$th largest element of ${\bf q,}$ and let $\left| \phi
_{k}^{\downarrow A}\right\rangle $ and $\left| \chi _{k}^{\downarrow
A}\right\rangle $ denote the vectors associated with $q_{k}^{\downarrow }.$

Now, identify every vector in the list $\{\left| \phi _{k}^{\downarrow
A}\right\rangle \}_{k=1}^{m}$ that cannot be obtained as a linear
combination of vectors with lower indices from this list. Suppose there a
number $m^{\prime }$ of such vectors, corresponding to a particular subset $%
S $ of the indices $\{1,2,...,m\},$ so that the set of vectors is denoted by 
$\{\left| \phi _{k}^{\downarrow A}\right\rangle \}_{k\in S}$. By definition,
this is a linearly independent set. The remaining vectors are denoted by $%
\{\left| \phi _{k}^{\downarrow A}\right\rangle \}_{k\in \bar{S}},$ where $%
\bar{S}$ is the set of indices that remain after removing the elements of $S$
from $\{1,2,...,m\}.$ Obviously the elements of $\{\left| \phi
_{k}^{\downarrow A}\right\rangle \}_{k\in \bar{S}}$ can all be written as
linear combinations of the elements of $\{\left| \phi _{k^{\prime
}}^{\downarrow A}\right\rangle \}_{k^{\prime }\in S,\text{ }k^{\prime }<k}.$
Finally, for future reference, we define $g(k)$ as the number of indices $%
k^{\prime }$ in $S$ such that $k^{\prime }\le k.$ It is clear from the
definition of $S$ that $g(k)\le k.$

Let $\{\left| \mu _{j}^{A}\right\rangle \}_{j=1}^{m^{\prime }}$ be the
ordered set of orthogonal vectors that are obtained by applying the
Gram-Schmidt orthogonalization procedure to $\{\left| \phi _{k}^{\downarrow
A}\right\rangle \}_{k\in S},$ in order of ascending $k.$ This new set yields
an $A$-orthogonal decomposition of $\left| \psi \right\rangle ,$ $D_{\text{%
A-orth}}=\{(c_{j},\left| \mu _{j}^{A}\right\rangle \otimes \left| \nu
_{j}^{B}\right\rangle )\}_{j=1}^{m^{\prime }},$ where $\left| \nu
_{j}^{B}\right\rangle =\left\langle \mu _{j}^{A}|\psi \right\rangle /c_{j}$
and $c_{j}=\left| \left\langle \mu _{j}^{A}|\psi \right\rangle \right| .$ It
also defines an $m$-element probability distribution ${\bf p=(}%
p_{1},p_{2},...,p_{m})$ where{\bf \ }$p_{j}\equiv \left| c_{j}\right| ^{2}$
for $j$ in the range $\{1,...,m^{\prime }\},$ and $p_{j}\equiv 0$ for $j$ in
the range $\{m^{\prime }+1,...,m\}.$

Let $P_{\phi }^{A}$ denote the projector onto the ray spanned by $\left|
\phi ^{A}\right\rangle $ and for convenience define $P_{\mu _{j}}^{A}\equiv
P_{\text{null}}$ for $j$ in the range $\{m^{\prime },...,m\}.$ The nature of
the Gram-Schmidt orthogonalization procedure ensures that for every $k\in S,$
$\left| \phi _{k}^{\downarrow A}\right\rangle =\sum_{j=1}^{g(k)}f_{j}\left|
\mu _{j}^{A}\right\rangle $ for some set of complex amplitudes \{$%
f_{j}\}_{j=1}^{g(k)}$. Thus, $(\sum_{j=1}^{g(k)}P_{\mu _{j}}^{A})\left| \phi
_{k}^{\downarrow A}\right\rangle =\left| \phi _{k}^{\downarrow
A}\right\rangle ,$ or equivalently, $\sum_{j=1}^{g(k)}P_{\mu _{j}}^{A}\ge
P_{\phi _{k}}^{\downarrow A}.$ Since $g(k)\le k,$ this is trivially extended
to $\sum_{j=1}^{k}P_{\mu _{j}}^{A}\ge P_{\phi _{k}}^{\downarrow A}.$
Moreover, if $k\in \bar{S},$ then $\left| \phi _{k}^{\downarrow
A}\right\rangle $ can be written as a linear combination of the elements of $%
\{\left| \phi _{k^{\prime }}^{\downarrow A}\right\rangle \}_{k^{\prime }\in
S,\text{ }k^{\prime }<k},$ so that $\left| \phi _{k}^{\downarrow
A}\right\rangle =\sum_{j=1}^{g((h(k))}\bar{f}_{j}\left| \mu
_{j}^{A}\right\rangle $ for some set of complex amplitudes \{$\bar{f}%
_{j}\}_{j=1}^{g(h(k))}$, where $h(k)=\max_{k^{\prime }\in S,k^{\prime
}<k}k^{\prime }.$ It follows that $\sum_{j=1}^{g((h(k))}P_{\mu _{j}}^{A}\ge
P_{\phi _{k}}^{\downarrow A}$ for all $k\in \bar{S}$. Since $g(h(k))<k,$
this is trivially extended to $\sum_{j=1}^{k}P_{\mu _{j}}^{A}\ge P_{\phi
_{k}}^{\downarrow A}.$ It follows therefore that for every $k$ in the range $%
\{1,...,m\}$ we have $\sum_{j=1}^{k}P_{\mu _{j}}^{A}\ge P_{\phi
_{k}}^{\downarrow A}.$ Now, since $I^{B}\ge P_{\chi _{l}}^{\downarrow B},$
we can infer that $\sum_{j=1}^{k}P_{\mu _{j}}^{A}\otimes I^{B}\ge P_{\phi
_{k}}^{\downarrow A}\otimes $ $P_{\chi _{k}}^{\downarrow B},$ and by the
orthogonality of the projectors in the set $\{P_{\phi _{l}}^{\downarrow
A}\otimes P_{\chi _{l}}^{\downarrow B}\}_{l=1}^{k},$ we conclude that $%
\sum_{j=1}^{k}P_{\mu _{j}}^{A}\otimes I^{B}\ge \sum_{l=1}^{k}P_{\phi
l}^{\downarrow A}\otimes $ $P_{\chi _{l}}^{\downarrow B}$ for every $k$ in
the range $\{1,...,m\}.$

Now we note that the probability distributions ${\bf q}$ and ${\bf p}$ are
related to the projectors by $q_{l}^{\downarrow }=\left\langle \psi \right|
P_{\phi _{l}}^{\downarrow A}\otimes P_{\chi _{l}}^{\downarrow B}\left| \psi
\right\rangle $ and $p_{j}=\left\langle \psi \right| P_{\mu _{j}}^{A}\otimes
I^{B}\left| \psi \right\rangle .$ From the inequality derived above together
with lemma A.4, we find that $\sum_{j=1}^{k}p_{j}\ge
\sum_{l=1}^{k}q_{l}^{\downarrow }$ for every $k$ in the range $\{1,...,m\}.$
By lemma A.3, it follows that ${\bf p}$ majorizes ${\bf q.}$ {\bf QED.}

Finally, we prove lemma A.2.

{\bf Proof of lemma A.2}. An arbitrary $A$-orthogonal decomposition of $%
\left| \psi \right\rangle $ has the form $D_{A\text{-orth}}=\{(c_{j},\left|
\mu _{j}^{A}\right\rangle \otimes \left| \nu _{j}^{B}\right\rangle
)\}_{j=1}^{m},$ where the vectors $\{\left| \mu _{j}^{A}\right\rangle
\}_{j=1}^{m}$ are orthogonal, but $\{\left| \nu _{j}^{B}\right\rangle
\}_{j=1}^{m}$ need not be orthogonal nor even linearly independent. A
bi-orthogonal decomposition of $\left| \psi \right\rangle $ has the form $D_{%
\text{bi-orth}}=\{(\tilde{c}_{j},\left| \tilde{\mu}_{j}^{A}\right\rangle
\otimes \left| \tilde{\nu}_{j}^{B}\right\rangle )\}_{j=1}^{\tilde{m}},$
where both the vectors $\{\left| \tilde{\mu}_{j}^{A}\right\rangle \}_{j=1}^{%
\tilde{m}}$ and $\{\left| \tilde{\nu}_{j}^{A}\right\rangle \}_{j=1}^{\tilde{m%
}}$ form orthogonal sets. The probability distributions associated with each
decomposition are $(\left| c_{1}\right| ^{2},\left| c_{2}\right|
^{2},...,\left| c_{m}\right| ^{2})$ and $(\left| \tilde{c}_{1}\right|
^{2},\left| \tilde{c}_{2}\right| ^{2},...,\left| \tilde{c}_{\tilde{m}%
}\right| ^{2})$ respectively (even if there is more than one bi-orthogonal
decomposition for a particular state vector, these do not differ in their
coefficients). For ease of comparison of these distributions, we add zeroes
until the number of elements in each is equal to the dimensionality, $d,$ of
the Hilbert space ${\cal H}^{A}$. Denote the resulting distributions by $%
{\bf p}$ and ${\bf \tilde{p}}$ respectively. We establish that $S_{\left|
\psi \right\rangle }(D_{\text{bi-orth}})\le S_{\left| \psi \right\rangle
}(D_{A-\text{orth}})$ by showing that ${\bf \tilde{p}}$ majorizes ${\bf p.}$

To begin, we express the probabilities as expectation values of projectors.
We introduce an arbitrary orthogonal set of vectors $\{\left| \mu
_{j}^{A}\right\rangle \}_{j=m+1}^{d}$ which together with $\{\left| \mu
_{j}^{A}\right\rangle \}_{j=1}^{m}$ form an orthogonal basis for the Hilbert
space ${\cal H}^{A}$, and similarly for $\{\left| \tilde{\mu}%
_{j}^{A}\right\rangle \}_{j=1}^{\tilde{m}}.$ Then, we have for all $j$ in
the range $\{1,...,d\},$%
\begin{eqnarray*}
p_{j} &=&Tr_{A}(\rho ^{A}P_{\mu _{j}}^{A}),\text{ and} \\
\tilde{p}_{j} &=&Tr_{A}(\rho ^{A}P_{\tilde{\mu}_{j}}^{A}).
\end{eqnarray*}
Let the unitary operator that transforms the elements of $\{\left| \tilde{\mu%
}_{j}^{A}\right\rangle \}_{j=1}^{d}$ to the elements of $\{\left| \mu
_{j}^{A}\right\rangle \}_{j=1}^{d}$ be denoted by $U^{A},$ so that 
\[
\left| \mu _{j}^{A}\right\rangle =U^{A}\left| \tilde{\mu}_{j}^{A}\right%
\rangle . 
\]
It follows that 
\begin{eqnarray*}
p_{j} &=&Tr_{A}(\rho ^{A}U^{A\dag }P_{\tilde{\mu}_{j}}^{A}U^{A}) \\
&=&Tr_{A}(U^{A}\rho ^{A}U^{A\dag }P_{\tilde{\mu}_{j}}^{A}),
\end{eqnarray*}
where in the last step we have used the cyclic property of the trace. What
distinguishes the bi-orthogonal decomposition from other $A$-orthogonal
decompositions is that the projectors $\{P_{\tilde{\mu}_{k}}^{A}\}_{k=1}^{d}$
diagonalize $\rho ^{A}$, 
\[
\rho ^{A}=\sum_{k=1}^{d}\tilde{p}_{k}P_{\tilde{\mu}_{k}}^{A}. 
\]
Plugging this form of $\rho ^{A}$ into the expression for $p_{j},$ we obtain 
\[
p_{j}=\sum_{k=1}^{d}\left| U_{jk}^{A}\right| ^{2}\tilde{p}_{k}, 
\]
where $U_{jk}^{A}=\left\langle \tilde{\mu}_{j}\right| U^{A}\left| \tilde{\mu}%
_{k}\right\rangle .$ \strut By the unitarity of $U^{A}$, we find that $%
\sum_{j}\left| U_{jk}^{A}\right| ^{2}=\left\langle \tilde{\mu}_{k}\right|
U^{A\dag }U^{A}\left| \tilde{\mu}_{k}\right\rangle =1,$ and $\sum_{k}\left|
U_{jk}^{A}\right| ^{2}=\left\langle \tilde{\mu}_{j}\right| U^{A}U^{A\dag
}\left| \tilde{\mu}_{j}\right\rangle =1.$ Thus, the transition matrix
between the probability distributions ${\bf \tilde{p}}$ and ${\bf p}$ is
doubly stochastic, from which it follows by a well-known result (theorem
II.1.9 of Ref. \cite{Bhatia}) that ${\bf \tilde{p}}$ majorizes ${\bf p.}$ 
{\bf QED.}\newpage


\begin{references}
\bibitem{see Clifton}  This is consistent with the characterization found in
Ref. \cite{Cliftonproperties}
\end{references}
\end{document}